\documentclass[aps, twocolumn, nobibnotes, superscriptaddress, notitlepage, noeprint, prb]{revtex4-2}
\usepackage[usenames,dvipsnames]{xcolor}
\usepackage[linktoc=page,colorlinks,urlcolor=RedViolet,citecolor=RedViolet,linkcolor=RedViolet]{hyperref}
\usepackage{float}
\usepackage{graphicx}
\usepackage{mathrsfs}
\usepackage{overpic}
\usepackage{wrapfig}
\usepackage{braket}
\usepackage{color}
\usepackage{verbatim}
\usepackage{amssymb,amsmath,mathtools}
\usepackage{upgreek}
\usepackage{bbold}
\usepackage{cancel}
\usepackage{textgreek}
\usepackage[normalem]{ulem}

\definecolor{darkblue}{rgb}{0.0 0.0 0.78}
\definecolor{darkred}{rgb}{0.5 0.0 0.0}

\newcommand{\UMDphy}{Department of Physics, University of Maryland, College Park, Maryland 20742, USA}
\newcommand{\QTC}{Quantum Technology Center, University of Maryland, College Park, Maryland 20742, USA}
\newcommand{\UMDEECS}{Department of Electrical Engineering and Computer Science,
University of Maryland, College Park, Maryland 20742, USA}

\newcommand{\LincolnLab}{Lincoln Laboratory, Massachusetts Institute of Technology, Lexington, Massachusetts 02421, USA}
\newcommand{\ICpostdoc}{Intelligence Community Postdoctoral Research Fellowship Program, University of Maryland, College Park, Maryland 20742, USA}

\begin{document}

\title{Ramsey Envelope Modulation in NV Diamond Magnetometry}
\date{\today}

\author{Jner Tzern Oon}
%\thanks{These authors contributed equally to this work.}
%\email{jjoon@umd.edu}
\email[These authors contributed equally to this work. ]{jjoon@umd.edu}
\affiliation{\UMDphy}
\affiliation{\QTC}

\author{Jiashen Tang}
\email[These authors contributed equally to this work. ]{jjoon@umd.edu}
%\thanks{These authors contributed equally to this work.}
\affiliation{\UMDphy}
\affiliation{\QTC}

\author{Connor A. Hart}
\affiliation{\QTC}
\affiliation{\UMDEECS}

\author{Kevin S. Olsson}
\affiliation{\QTC}
\affiliation{\UMDEECS}
\affiliation{\ICpostdoc}

\author{Matthew J. Turner}
\affiliation{\QTC}
\affiliation{\UMDEECS}

\author{Jennifer M. Schloss}
\affiliation{\LincolnLab}

\author{Ronald L. Walsworth}
%\email{walsworth@umd.edu}
\affiliation{\UMDphy}
\affiliation{\QTC}
\affiliation{\UMDEECS}

\begin{abstract}
Nitrogen-vacancy (NV) spin ensembles in diamond provide an advanced magnetic sensing platform, with applications in both the physical and life sciences. 
The development of isotopically engineered $^{15}$NV diamond offers advantages over naturally occurring $^{14}$NV for magnetometry, due to its simpler hyperfine structure. However, for sensing modalities requiring a bias magnetic field not aligned with the sensing NV axis, the absence of a quadrupole moment in the $^{15}$N nuclear spin leads to pronounced envelope modulation effects in time-dependent measurements of $^{15}$NV spin evolution. While such behavior in spin echo experiments are well studied, analogous effects in Ramsey measurements and the implications for magnetometry remain under-explored. 
Here, we derive the modulated $^{15}$NV Ramsey response to a misaligned bias field, using a simple vector description of the effective magnetic field on the nuclear spin. The predicted modulation properties are then compared to experimental results, revealing significant magnetic sensitivity loss if unaddressed. We demonstrate that double-quantum coherences of the NV $S=1$ electronic spin states %can 
dramatically suppress these envelope modulations, while additionally proving resilient to other parasitic effects such as strain heterogeneity and temperature shifts.
\end{abstract}

\maketitle

\section{Introduction}

\begin{figure}[t]
    \centering
    \includegraphics[width=\linewidth]{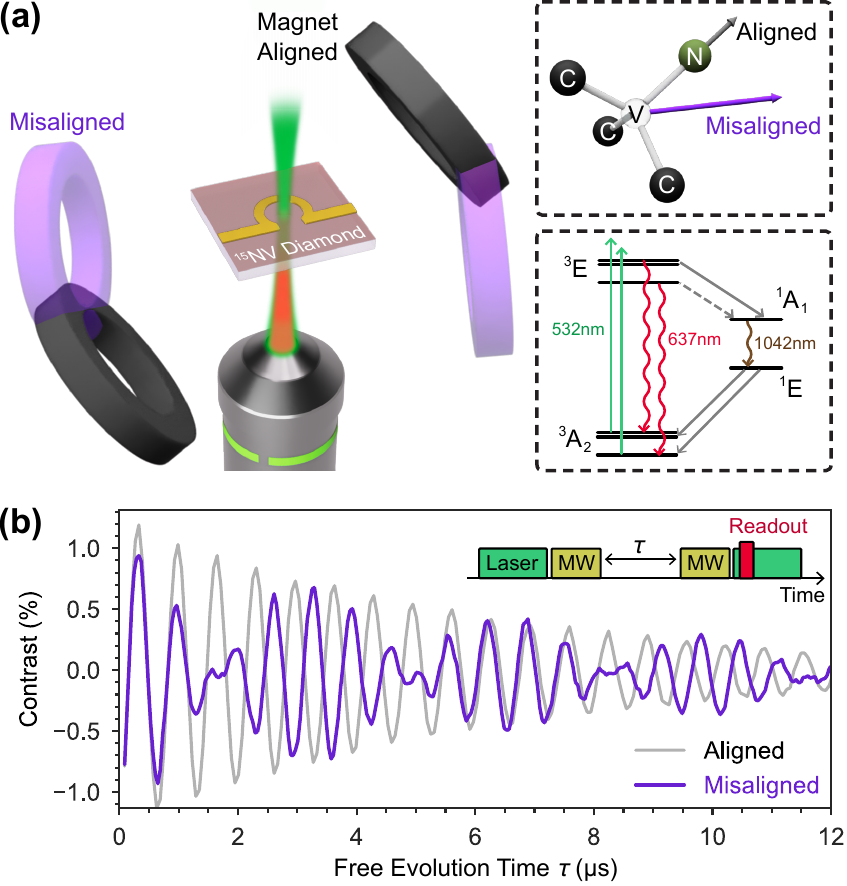}
    \caption{(a) Experimental setup schematic. Diamond containing an ensemble ($\approx0.3$\,ppm) of $^{15}$NV centers is probed using 532\,nm green laser excitation through a microscope objective, with red fluorescence collected back along the same path. A gold $\Omega$-shaped waveguide is placed above the NV layer for microwave (MW) delivery.
    A pair of ring magnets apply a bias magnetic field at two distinct orientations, aligned and misaligned with the symmetry axis for one class of NV centers. Top right inset: diamond crystal axes and bias field orientations relative to a single NV quantization axis. Bottom right inset: NV electronic energy level diagram of ground and excited states. (b) Example Ramsey time series measured for bias fields aligned (gray) and misaligned (blue) to one class of NV centers. Effect of electron Ramsey envelope modulation (EREEM) is clearly observable for data with the misaligned field. Upper inset: Ramsey pulse sequence diagram.} 
    \label{fig:1}
\end{figure}

Ensembles of negatively charged nitrogen-vacancy (NV) centers in diamond are a leading quantum sensing platform, particularly for applications in magnetometry.
The NV center has a magnetically sensitive electronic triplet ground state with spin $S=1$ that can be optically initialized and read out, and coherently manipulated using microwave fields, while operating at ambient conditions.
%Additionally, the NV electronic spin is coupled via a hyperfine interaction to the constituent nitrogen nuclear spin, with spin $I=1$ and $I=1/2$ for the stable $^{14}$N and $^{15}$N isotopes, respectively.  

Demonstrations of sensing or imaging of static and broadband (DC) magnetic fields have predominantly used continuous-wave optically detected magnetic resonance (CW-ODMR) techniques. However, the achievable volume-normalized magnetic sensitivity in CW-ODMR is constrained by competing effects of the optical and microwave fields applied during sensing \cite{Barry2020}. Alternatively, pulsed measurement protocols such as Ramsey interference magnetometry can be employed to measure DC magnetic fields \cite{Ramsey1950, Rondin2014}. By separating spin control and readout from the sensing interval, pulsed measurements enable the use of increased optical and microwave intensities to improve sensitivity. As a result, Ramsey protocols have produced some of the best volume-normalized DC sensitivities reported to date for NV ensembles \cite{Balasubramanian2019, Hart2021}. 
 
Beyond advancements in sensing protocols, the optimization of diamond material properties provides a crucial path towards improvements in volume-normalized sensitivity. 
In particular, $^{15}$NV centers found in $^{15}$N-enriched diamond provide practical advantages over the naturally abundant $^{14}$NV due to the nuclear spin $I=1/2$ of $^{15}$N, as compared to $I=1$ for $^{14}$N. 
For sensing, this difference translates to increased signal contrast during optical readout while driving a single $^{15}$NV hyperfine resonance, as the nuclear spin population is only distributed between two states.
%In particular, $^{15}$NV centers found in $^{15}$N-enriched diamond provide practical advantages over the naturally abundant $^{14}$NV due to the simplified nuclear spin $I=1/2$ property of $^{15}$N, as compared to $I=1$ for $^{14}$N. For sensing, this translates to increased signal contrast during optical readout while driving a single hyperfine resonance, as the nuclear spin population is only distributed between two nuclear states for $^{15}NV$
%In particular, $^{15}$NV centers found in $^{15}$N-enriched diamond provide practical advantages over the naturally abundant $^{14}$NV, as the nuclear spin $I=1/2$ for $^{15}$N produces a distribution of spin population between only two nuclear states, compared to three states for the $^{14}$N isotope $(I=1)$. 
%For sensing, this difference translates to increased signal contrast during optical readout while driving a single hyperfine resonance.
%By tuning the frequency of an applied microwave field to the midpoint of the hyperfine-split resonances, common Rabi nutation rate.
Both $^{15}$NV hyperfine-split electronic resonances can also be driven simultaneously with the same Rabi nutation rate by tuning the frequency of the applied microwave field to the midpoint of the splitting, enabling more uniform spin control. In addition, the two-level nuclear spin system simplifies quantum logic protocols that exploit the coupled electron-nuclear system for enhanced sensing \cite{Lovchinsky2016, Arunkumar2022}.

 % sorry for the comment bombs how about this? We don't need to specify improved readout fidelity? JJ: Rearranged slightly but looks good!
% In addition, the two-level nuclear spin system simplifies quantum logic enhanced sensing protocols that exploit the coupled electron-nuclear system \cite{Lovchinsky2016, Arunkumar2022}.
%For sensing, this difference translates to increased contrast and reduced hyperfine broadening for $^{15}$NV. It also enables both hyperfine-split resonances to be driven simultaneously with the same Rabi nutation rate, by tuning the frequency of the applied microwave field to the midpoint of the splitting.
%For sensing, this difference translates to increased signal contrast and reduced hyperfine broadening for $^{15}$NV. Compared to the triplet splitting for $^{14}$NV, the pair of hyperfine-split resonances for $^{15}$NV also enables more uniform spin control by equally detuning the frequency of the applied microwave field between the pair of resonances, resulting in a common Rabi nutation rate.

%The ease with which Ramsey magnetometry can be implemented with $^{15}$N-enriched diamond depends on the bias magnetic field commonly employed for magnetic sensing to break the ms = \pm 1 ground state spin degeneracy. In practice, the field magnitude and orientation is often constrained by the desired sensing modality or system to be studied.

The ease with which Ramsey magnetometry can be implemented with $^{15}$N-enriched diamond depends on the bias magnetic field commonly applied to break the ground state electronic spin degeneracy, associated with spin sublevels $m_s=\pm1$. In practice, the bias field magnitude and orientation is often constrained by the desired sensing modality or system to be studied.
%A bias magnetic field is commonly employed for magnetic sensing to break the $m_s=\pm1$ ground state spin degeneracy.
%To break the NV ground state spin degeneracy for the $m_s=\pm1$ energy levels at zero field, a bias magnetic field is commonly applied for sensing experiments. 
%The ease with which Ramsey magnetometry can be implemented with $^{15}$N-enriched diamond depends on the bias magnetic field orientation and magnitude, often constrained by the desired sensing modality or system to be studied. 
For example, full vector reconstruction of magnetic fields in three dimensions typically requires a bias field
oriented to produce a unique projection onto each class of NV centers across the four crystal axes \cite{Glenn2017, Schloss2018, Turner2020, Levine2019, Turner2020}. This approach ensures that the resonances associated with each class of NV centers are non-degenerate and individually addressable with microwave control. Alternatively, the bias field may be applied to spectrally overlap two or more NV classes to increase the number of spins participating in sensing, improving sensitivity \cite{Jensen2014, Barry2016, Levine2019, Lenz2021, Alsid2022inprep}.

In the presence of such misaligned fields (Fig.~\ref{fig:1}), we observe envelope modulations in $^{15}$NV Ramsey measurements,
which negatively impacts sensitivity if left unaddressed. The physical origin of this behavior can be attributed to the electron-nuclear hyperfine coupling of the $^{15}$NV center in the presence of a transverse magnetic field. This effect resembles the well-known electron spin echo envelope modulation (ESEEM), which has received extensive study for over half a century in NMR systems and more recently in solid state defects \cite{Rowan1965, Gaebel2006, Childress2006}. However, the analogous effect on a Ramsey measurement and the resulting impact on NV magnetic sensing has yet to be detailed.

In this paper, we characterize this effect, which we refer to as electron Ramsey envelope modulation or EREEM. First, we model the Ramsey envelope properties by considering an NV electronic spin coupled to its native nitrogen nuclear spin. The resulting EREEM predictions are compared to experimental results, showing good agreement for $^{15}$NV ensembles across a range of magnetic field magnitudes and misalignments. We then discuss the impact of EREEM on NV-diamond magnetic sensitivity, considering typical operating conditions used for magnetometry. Finally we study EREEM in the context of double-quantum (DQ) protocols, which leverage superpositions of NV electronic spin states $\ket{m_s = \pm1}$ for magnetometry. We demonstrate dramatic suppression of envelope modulations in DQ Ramsey measurements. These results provide further motivation for the use of DQ sensing schemes, in addition to their documented robustness to strain gradients and temperature drift. \cite{Reinhard2012, Fang2013, Mamin2014, Bauch2018, Hart2021}.

\section{Electron Ramsey Envelope Modulation (EREEM)}
\label{sec:theory}

\begin{figure}[t]
    \centering
    \includegraphics[width=\linewidth]{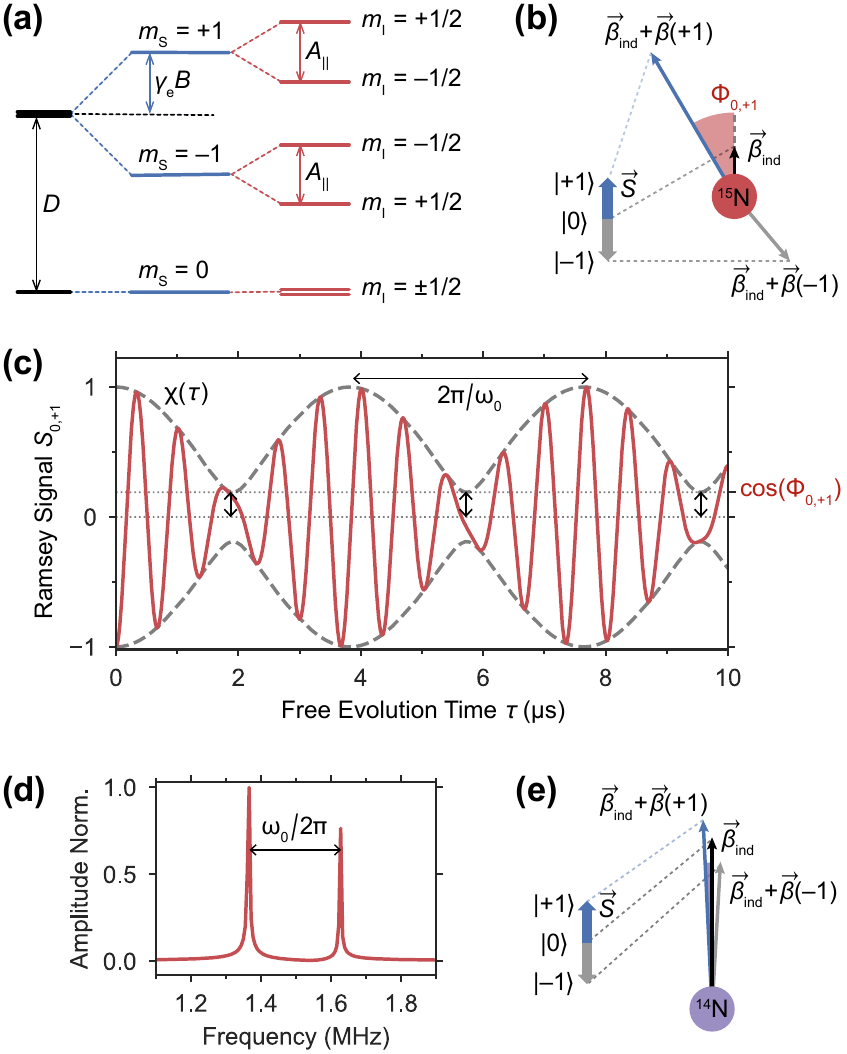}
    \caption{Energy level diagram, 
    modulated $^{15}$NV Ramsey free induction signal (EREEM), and vector models of the effective magnetic fields for nuclear spins $^{15}$N and $^{14}$N. (a) Energy level diagram for the $^{15}$NV ground state, with splittings due to Zeeman and hyperfine interactions assuming an aligned bias field. (b) Effective magnetic field vectors on the $^{15}$N nuclear spin due to a misaligned bias magnetic field, for each electronic spin state $\ket{m_s}$. Each effective nuclear field consists of an electronic-spin-dependent contribution, $\vec{\beta}(m_s)$, and a component $\vec{\beta}_\text{ind}$ independent of the electronic spin state.
    (c) Example $^{15}$NV Ramsey free induction signal due to a 15$^\circ$ misaligned bias magnetic field, showing envelope modulation. 
    The amplitude modulation $\chi(\tau)$ is indicated, showing oscillations between $\chi_\text{min}=\cos{\left(\Phi_{0,+1}\right)}$ and 1, with a characteristic period $2\pi/\omega_0$.
    (d) Power spectrum of the Ramsey signal from (c). (e) Effective magnetic field vectors on the $^{14}$N nuclear spin. For all three electronic spin states, the effective nuclear fields are nearly parallel due to a dominant nuclear quadrupolar field.}
    \label{fig:2}
\end{figure}

This section presents a derivation of EREEM properties, described by a simple vector model of the effective magnetic field on the nitrogen nuclear spin - which, importantly, is dependent on the NV electronic spin state. First, the $^{15}$NV center is modeled by an electronic spin system ($S=1$) coupled to the native $^{15}$N nuclear spin ($I=1/2$). Under the application of a bias magnetic field $\vec{B}=(B_x, B_y, B_z)$, the ground state Hamiltonian $H$ can be written as \cite{Doherty2013}
\begin{equation}
\label{eq:HGSlab}
\frac{H}{\hbar} = D\hat{S}_z^2 - \gamma_e \vec{B}\cdot\vec{S} - \gamma_n \vec{B}\cdot\vec{I} + \vec{S}\cdot \mathbf{A}\cdot\vec{I}.
\end{equation}
The vectors $\vec{S}=(\hat{S}_x, \hat{S}_y, \hat{S}_z)$ and $\vec{I}=(\hat{I}_x, \hat{I}_y, \hat{I}_z)$ contain the electronic and nuclear spin operators, respectively, with corresponding gyromagnetic ratios  $\gamma_e=2\pi\times-2.8024$\,MHz/G and $\gamma_n=2\pi\times -431.6$\,Hz/G. The room temperature zero field splitting  $D \approx 2\pi \times 2.87$\,GHz \cite{Acosta2010} sets the electron quantization axis along $\hat{z}$. The NV hyperfine interaction is described by the diagonal tensor $\mathbf{A} = \left(\begin{smallmatrix}
A_\perp & 0 & 0\\
0 & A_\perp & 0\\
0 & 0 & A_{||}
\end{smallmatrix}\right)$,  
with transverse and longitudinal components $A_\perp=2\pi\times3.65$\,MHz and $A_{||}=2\pi\times3.03$\,MHz, respectively \cite{Felton2009}. This ground state energy level structure is depicted in Fig.~\ref{fig:2}(a), for a magnetic field of magnitude $B$ aligned along the $\hat{z}$ direction.

The $C_{3v}$ symmetry of the NV center allows us to restrict the magnetic field to the $x$-$z$ plane without any loss in generality \cite{myers2016quantum}. 
For this study, we consider bias magnetic fields $B<200$\,G. Within this field regime, the zero field splitting term $D \hat{S}_z^2$ sets the dominant energy scale in the Hamiltonian, allowing us to
%For the bias magnetic field magnitudes considered in this study, $B\lesssim110$\,G, the zero field splitting term $D \hat{S}_z^2$ sets the dominant energy scale in the Hamiltonian. 
treat contributions not commuting with $S_z$ as non-secular perturbations. Accurate to second order in perturbation theory, a leading order correction to the secular Hamiltonian can be obtained \cite{Childress2006, myers2016quantum, SI}.
After transforming into a frame resonant with the two electronic transitions $m_s=0\leftrightarrow\!+1$ and $0\leftrightarrow\!-1$ \cite{Mamin2014}, the following Hamiltonian under the rotating wave approximation is found:
 \begin{equation}
 \label{eq:Hrot}
\begin{aligned}
\frac{\widetilde{H}}{\hbar} = &A_{||} \hat{S}_z \hat{I}_z -\gamma_n B_z \hat{I}_z \\ &- (1-2\kappa) \gamma_n B_x \hat{I}_x - 3\kappa \gamma_n B_x \hat{S}_z^2 \hat{I}_x.
\end{aligned}
\end{equation}
Here, a dimensionless factor $\kappa \equiv \frac{\gamma_e A_\perp}{\gamma_n D}\approx 8.26$ describes an effective amplification of the bare nuclear spin response to a transverse magnetic field $B_x$, by a factor of $1-2\kappa\approx -15.5$.

The contributions to Eq.~\eqref{eq:Hrot} can be separated into two categories. The first category consists of terms that depend on $\hat{S}_z$ or equivalently $m_s$. The sum of these terms can be described as an effective vector magnetic field $\vec{\beta}(m_s)$ on the nuclear spin.
The terms that do not contain $\hat{S}_z$ can be represented by a spin-independent effective field $\vec{\beta}_\text{ind}$, with a constant coupling to the nuclear spin regardless of the electronic spin state $m_s$. The resulting Hamiltonian can thus be summarized as
\begin{equation}
\label{eq:Hnuc}
\begin{aligned}
\frac{\widetilde{H}_n}{\hbar}
 &= -\gamma_n
\left(\vec{\beta}_{\text{ind}} + \vec{\beta}(m_s)\right)
\cdot \vec{I}.
\end{aligned}
\end{equation}
For a given electronic spin state $\ket{m_s}$, the nuclear spin precesses around an effective magnetic field $\vec{\beta}_{\text{ind}} + \vec{\beta}(m_s)$ with a Larmor frequency
$\omega_{m_s} = \big| \gamma_n \big(\vec{\beta}_{\text{ind}} + \vec{\beta}(m_s)\big)\big|$. For $m_s=0$ in particular, there are no terms in Eq.~\eqref{eq:Hrot} that depend on the electronic spin state, such that $\vec{\beta}(0)=0$ and $\omega_{0} = \big|\gamma_n \beta_{\text{ind}}\big|$.
These field vectors are visualized in Figure~\ref{fig:2}(b). For simplicity, the coordinate system is rotated so the spin-independent field $\vec{\beta}_{\text{ind}} = \beta_{\text{ind}} \hat{z}'$ now lies along the newly defined $z'$-axis \cite{SI}. In this frame, the angle between $\vec{\beta}_{\text{ind}}$ and an effective field $\vec{\beta}_{\text{ind}}+\vec{\beta}(m_s)$ is given by
\begin{equation}
\begin{aligned}
\label{eq:phims}
\phi_{m_s} = \tan^{-1}{\left(\frac{\beta_{x'}(m_s)}{\beta_{\text{ind}} + \beta_{z'}(m_s)}\right)},
\end{aligned}
\end{equation}
where $\beta_{z'}(m_s)$ and $\beta_{x'}(m_s)$ denote components of $\vec{\beta}(m_s)$ parallel and perpendicular to $\hat{z'}$, respectively.
For distinct electronic spin states $\ket{i}$ and $\ket{j}$, we define the angle between the two corresponding nuclear fields as $\Phi_{i,j}$, with the example $\Phi_{0,+1}=\phi_{+1}$ shown in Fig.~\ref{fig:2}(b). 

Using this vector description, we derive the expected Ramsey envelope modulation as a function of the free evolution time $\tau$. For an initial superposition of electronic spin states $\ket{i}$ and $\ket{j}$, the Ramsey signal $S_{i,j}(\tau)$, up to an overall phase, is given by
\begin{equation}
\label{eq:EREEMsig}
\begin{aligned}
S_{i,j}(\tau) = 
&\cos{\left(\Phi_{i,j}\right)}
 \sin{\left(\frac{\omega_i \tau}{2}\right)} \sin{\left(\frac{\omega_j \tau}{2}\right)}
 \\&-\cos{\left(\frac{\omega_i \tau}{2}\right)} \cos{\left(\frac{\omega_j \tau}{2}\right)}.
\end{aligned}
\end{equation}
Figure~\ref{fig:2}(c) shows an example simulated Ramsey response $S_{0,+1}(\tau)$ due to a single-quantum coherence between $m_s=0\leftrightarrow\!+1$, for a bias field of magnitude $B=100$\,G misaligned from the NV axis by an angle $\theta= 15^\circ$. The Ramsey signal oscillation is modulated by an envelope with a slow characteristic beat frequency $\omega_0$, i.e., an example of EREEM. The corresponding power spectrum is shown in Fig.~\ref{fig:2}(d), revealing two peaks centered around $\omega_{+1}/2\approx A_{||}/2=2\pi\times1.515$\,MHz, with a frequency splitting of magnitude $\omega_0$. 
%The corresponding power spectrum is shown in Fig.~\ref{fig:2}(d).
At any given time $\tau$, the maximum Ramsey signal contrast must be corrected by a multiplicative factor $\chi(\tau)$, due to this envelope modulation. This amplitude modulation factor $\chi(\tau)$ oscillates as a function of $\tau$ between values $\chi_\text{min} = |\cos{\left(\Phi_{0,+1}\right)}|$ and $\chi_\text{max}=1$, indicating points of minimum and maximum contrast, respectively. The depth of this modulation can be inferred from the angle $\Phi_{0,+1}$ between the participating effective nuclear fields. 

To connect this vector model to the expected EREEM behavior in experimentally realistic conditions, we first consider a magnetic field aligned with the NV axis.
Since $B_x=0$, the Hamiltonian in Eq.~\eqref{eq:Hrot} consists only of nuclear spin contributions along $\hat{I}_z$.
Consequently, the effective nuclear fields $\vec{\beta}_{\text{ind}}$ and $\vec{\beta}_{\text{ind}}+\vec{\beta}(+1)$ are parallel, such that $\Phi_{0,+1}=0$. No Ramsey envelope modulation should be observed, as the amplitude modulation factor remains constant: $\chi_\text{min} = |\cos{\left(\Phi_{0,+1}\right)}|= \chi_\text{max} = 1$.
However, in the presence of a magnetic field not aligned with the NV axis ($B_x\neq0$), the nuclear spin experiences an enhanced transverse magnetic field determined by the factor $\kappa$. The effective nuclear fields are no longer aligned, $\Phi_{0,+1}>0$, which should lead to observable envelope modulation. At bias magnetic fields where $\vec{\beta}_{\text{ind}}$ and $\vec{\beta}_{\text{ind}}+\vec{\beta}(+1)$ are orthogonal, $\chi_\text{min} = 0$ and the Ramsey signal contrast at modulation nodes is maximally suppressed.

This vector model can be readily extended to the $^{14}$NV center, with some modifications. Besides straightforward changes to the physical constants $A_\perp$, $A_{||}$, and $\gamma_n$, an additional nuclear quadrupolar interaction term $Q \hat{I}_z^2$ contributes to the Hamiltonian in Eq.~\eqref{eq:HGSlab}, with quadrupolar coupling constant $Q=2\pi\times-4.945$\,MHz~\cite{Smeltzer2009, Steiner2010}. This inclusion dramatically changes the Ramsey envelope properties, by contributing a large quantizing field of magnitude $|Q/\gamma_n| \approx 16\,000$\,G to $\vec{\beta}_{\text{ind}}$ \cite{Childress2006}.
In the small magnetic field regime $B \ll |Q/\gamma_n|$, the effective nuclear fields $\vec{\beta}_{\text{ind}}+\vec{\beta}(m_s)$ are dominated by the spin-independent contribution $\vec{\beta}_{\text{ind}}$, which is visualized in Fig.~\ref{fig:2}(e).
The small angle between the nuclear field vectors results in $\chi_\text{min}\approx1$, and suppressed EREEM for $^{14}$NV.

\section{Experimental Methods}
\label{sec:methods}
The measurements for this study utilize a custom-built microscope setup and a $100\,$\textmu m-thick, $^{15}$N-enriched CVD diamond layer (NV $T_2^*=5\,$\textmu s, $[\text{N}]\approx3$\,ppm, $>99.995\%$ $^{12}$C), grown by Element Six Ltd.~on a 2$\times$2$\times$0.5\,mm$^3$ high-purity diamond substrate, as shown in Fig.~\ref{fig:1}(a). Post-growth treatment via electron irradiation and annealing increases the NV concentration to approximately 0.3\,ppm. 

Initialization of the NV ensemble electronic spin states is accomplished via pulsed 532\,nm excitation, generated by a continuous-wave laser gated by an acoustic optical modulator (AOM). The light is focused onto the NV layer using a microscope objective, which is also used to route the outgoing NV fluorescence for spin state readout. Microwave pulses are synthesized by signal generators and controlled by switches for single- or double-quantum control of the NV spin states. The microwave drive fields are delivered through an $\Omega$-shaped planar waveguide, fabricated onto a sapphire substrate. The bias magnetic field is applied using two identical permanent ring magnets equally spaced from the diamond sample.
The field magnitude is manually adjusted by varying the separation between the magnets. To control the field misalignment angle from the target NV axis, two automated rotation stages are used to adjust the yaw and pitch of the magnet pair with a nominal accuracy of 0.047$^\circ$. The bias magnets provide a homogeneous field magnitude of up to $\sim$150\,G over an illumination spot size of $\sim$20\,\textmu m in diameter on the NV-rich layer. Additional details regarding the experimental setup are provided in the Supplementary Material \cite{SI}.
 
To accurately determine the bias field magnitude $B$ and misalignment angle $\theta$, pulsed optically detected magnetic resonance (pulsed-ODMR) spectroscopy is employed to probe the NV ground state spin resonances. 
First, the field is aligned to a single NV axis by adjusting the magnets until the pulsed-ODMR spectra of the other three misaligned NV classes overlap. Using weak microwave $\pi$-pulses of duration $\sim 1\,$\textmu s each \cite{Dreau2011}, the resonance spectrum of the aligned NV axis is recorded for both electronic transitions $m_s=0 \leftrightarrow\! +1$ and $m_s=0 \leftrightarrow\! -1$. These transitions are separated by $\Delta_{\pm1} = 2 \gamma_e B$, which is used to estimate the bias field magnitude $B$. 
The magnets are then rotated away from the NV quantization axis, and Ramsey measurements are performed at a range of misalignment angles $\theta$. At each position, the ODMR spectrum is again recorded to measure $\Delta_{\pm1}$. This frequency difference is determined by the field projection along the NV axis, $\Delta_{\pm1}=2\gamma_e B \cos{\theta}$, which is then used to estimate $\theta$. Additional details are provided in the Supplemental Material \cite{SI}.
Field misalignment angles of up to $\theta\approx 40^\circ$ can be accessed, limited by the geometrical constraints of the setup.

\section{MEASURED EREEM Properties}
\label{sec:envprops}
\begin{figure}[t]
    \centering
    \includegraphics[width=\linewidth]{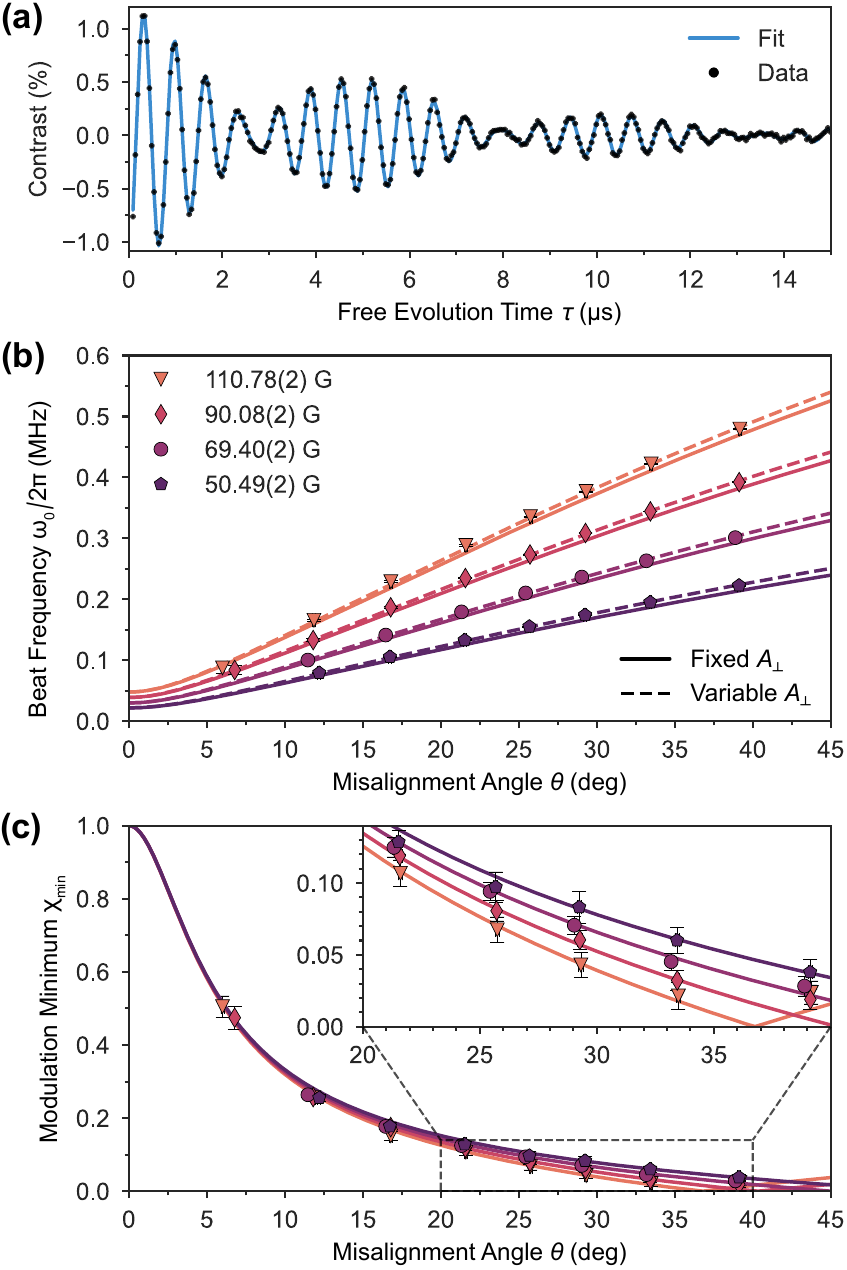}
    \caption{Ramsey experimental data and envelope properties from model fits. (a) Example Ramsey time series at field magnitude $B=90.08(2)$\,G and misalignment angle $\theta=16.79(9)^\circ$. Data is shown in black and the fit in blue. The observed slow beating is an example of EREEM. (b) Estimates of envelope beat frequency $\omega_0$ extracted from fits to experimental data, collected across a range of magnetic field magnitudes and orientations.
    Theoretical predictions for $\omega_0$ are shown in solid lines for each magnetic field. After allowing the transverse hyperfine constant $A_\perp$ to vary as a fitting parameter, updated $\omega_0$ predictions at each field are shown with dashed lines. (c) Estimates of the relative contrast at amplitude modulation nodes, $\chi_\text{min}$, for the magnetic field configurations used in (b). Solid lines indicate theoretical predictions with $A_\perp$ fixed. Inset: Magnified view of $\chi_\text{min}$ values for misalignment angles $20^\circ<\theta<40^\circ$.}
    \label{fig:3}
\end{figure}

To study the properties of electron Ramsey envelope modulation (EREEM), we perform a series of Ramsey experiments involving the electronic basis states $\ket{m_s=0}$ and $\ket{m_s=+1}$. The envelope properties are extracted using fits to the measured Ramsey time series, repeated at various magnetic field magnitudes and misalignment angles, with results shown in Fig.~\ref{fig:3}. 
The Ramsey protocol consists of two microwave $\pi/2$-pulses used to drive the $m_s=0\leftrightarrow\!+1$ transition, spaced by a variable free precession interval $\tau$. 
To avoid artifacts in the Ramsey fringes due to microwave detuning errors, the driving frequency is carefully calibrated to probe the center of the two hyperfine resonances \cite{SI}. This results in equal detunings of magnitude $A_{||}/2 =2\pi\times1.515$\,MHz from each hyperfine transition, giving the characteristic Ramsey fringe frequency. A slow beating (EREEM) is observed when exposed to a misaligned field, with an example shown in Fig \ref{fig:3}(a).

The measured Ramsey signals are fit to a modified form of Eq.~\eqref{eq:EREEMsig},  which incorporates an exponential decay $e^{-(\tau/T_2^*)^p}$ with a characteristic dephasing time $T_2^*$ and stretch factor $p$. 
To further adapt the expression to experimental data, an overall amplitude scaling factor,
a vertical offset, and phase offsets are all included in the fit function \cite{SI}. From the resulting fits, two frequencies $\omega_0$ and $\omega_{+1}$ are obtained. The values for the envelope beat frequency $\omega_0$ are plotted in Fig.~\ref{fig:3}(b), with 95\% confidence intervals indicated by error bars \cite{SI, DiCiccio1996}. For each magnetic field magnitude, theoretical predictions for $\omega_0$ are also plotted as solid curves, obtained from the spin-independent contributions to Eqs. \eqref{eq:Hrot} and \eqref{eq:Hnuc},
\begin{equation}
\label{eq:omega0}
\begin{aligned}
\omega_0
= \left|\gamma_n \beta_{\text{ind}}\right|
&=\left|\gamma_n\right|\sqrt{B_z^2 + (1-2\kappa)^2 B_x^2}\\
&=\left|\gamma_n\right| B\sqrt{1+ 4(\kappa^2-\kappa) \sin^2{\theta}}.
\end{aligned}
\end{equation}
As expected from Eq.~\eqref{eq:omega0}, the measured envelope beat frequency $\omega_0$ increases with both the magnetic field magnitude $B$ and misalignment angle $\theta$.  

Notably, small differences are observed between theoretical predictions of $\omega_0$ (solid curves in Fig. \ref{fig:3}(b)) and fits to experimental data, ranging from around 2\% to 6\% across the measurements presented in Fig.~\ref{fig:3}.
To explore this inconsistency, we first perform full-Hamiltonian numerical simulations of Ramsey spin dynamics and compare the observed envelope beat frequency to Eq.~\eqref{eq:omega0}, which was originally obtained using second order perturbation theory.
These simulations are conducted using the QuTiP package \cite{qutip1, qutip2} in Python. The NV system is described by the lab frame Hamiltonian from Eq.~\eqref{eq:HGSlab} and the pulse sequence is implemented using time-dependent AC magnetic field contributions. For the field configurations considered in Fig.~\ref{fig:3}, strong agreement is observed between Eq.~\eqref{eq:omega0} and the results of QuTiP simulations, with differences in $\omega_0 \lesssim 1\%$ (see the Supplemental Material \cite{SI}). These results, however, do not fully account for the observed discrepancies in $\omega_0$.
%Given this result, we do not consider the limitations of second order perturbation theory as the sole contribution to the observed discrepancy.

%Next, we consider the contributions to Eq.~\eqref{eq:omega0}

Interestingly, the agreement between analytical and experimental results is improved when the enhancement parameter $\kappa=\frac{\gamma_e A_\perp}{\gamma_n D}$ in Eq.~\eqref{eq:omega0} is allowed to deviate. 
%by close to 10\%.
Besides the transverse hyperfine constant $A_\perp$, other contributions to $\kappa$ include the well-established gyromagnetic ratios $\gamma_e$ and $\gamma_n$, and the zero field splitting $D$. We determine $D$ to be 2870.71(4)\,MHz using pulsed-ODMR measurements, consistent ($<\!0.03\%$ deviation) with the value assumed for analytical predictions and QuTiP numerical simulations.
%the zero field splitting $D$ and gyromagnetic ratios $\gamma_e$ and $\gamma_n$, all of which have well-established values. 
%Although temperature changes and crystal strain inhomogeneities can lead to small changes in $D$, these only account for less than
%, our measurements of $D$ using pulsed-ODMR techniques are consistent with the values used in these studies
%we do not expect thi the fractional change to $D$ 
%, which we determine to be $\lesssim 1$\,MHz across the interrogation volume
In contrast, an experimental determination of $A_\perp$ for $^{15}$NV has (to our knowledge) only been reported once, by Felton et al. \cite{Felton2009}, using EPR studies at higher fields $B\sim 2000$\,G. Given the simple relationship between the envelope beat frequency $\omega_0$ and $A_\perp$ at low fields (via $\kappa$ in Eq.~\eqref{eq:omega0}), EREEM presents a direct probe of $A_\perp$ in this regime.
With this in mind, we conduct a phenomenological fit of Eq.~\eqref{eq:omega0} to measurements of $\omega_0$ at each magnetic field, with $A_\perp$ as the sole degree of freedom. Using the adjusted $A_\perp$ values at each field, the corresponding values for $\omega_0$ from Eq.~\eqref{eq:omega0} are shown as dashed lines in Fig.~\ref{fig:3}(b). We obtain values of $A_\perp/2\pi$ between $3.74$\,MHz and $3.80$\,MHz across the fields considered here, differing slightly from the previously reported value of $3.65(3)$\,MHz by around $3\%$. 
We note that an observed deviation with respect to $B$ (see the Supplemental Material \cite{SI}) warrants a more thorough study across an extended range of magnetic fields, left as a subject for future work.

Figure~\ref{fig:3}(c) shows excellent agreement between theoretical predictions and experimentally observed values of the relative contrast at amplitude modulation nodes $\chi_\text{min}=|\cos{\left(\Phi_{0,+1}\right)}|$. 
Even for modest misalignments of $\sim 10^\circ$, the contrast at the nodes of the envelope modulation is reduced to around 30\% of its maximum value. The resulting implications for magnetic field sensitivity are discussed in the following section.

\section{Modulation Amplitude and Impact on Magnetometry}

\begin{figure}[t]
    \centering
    \includegraphics[width=\linewidth]{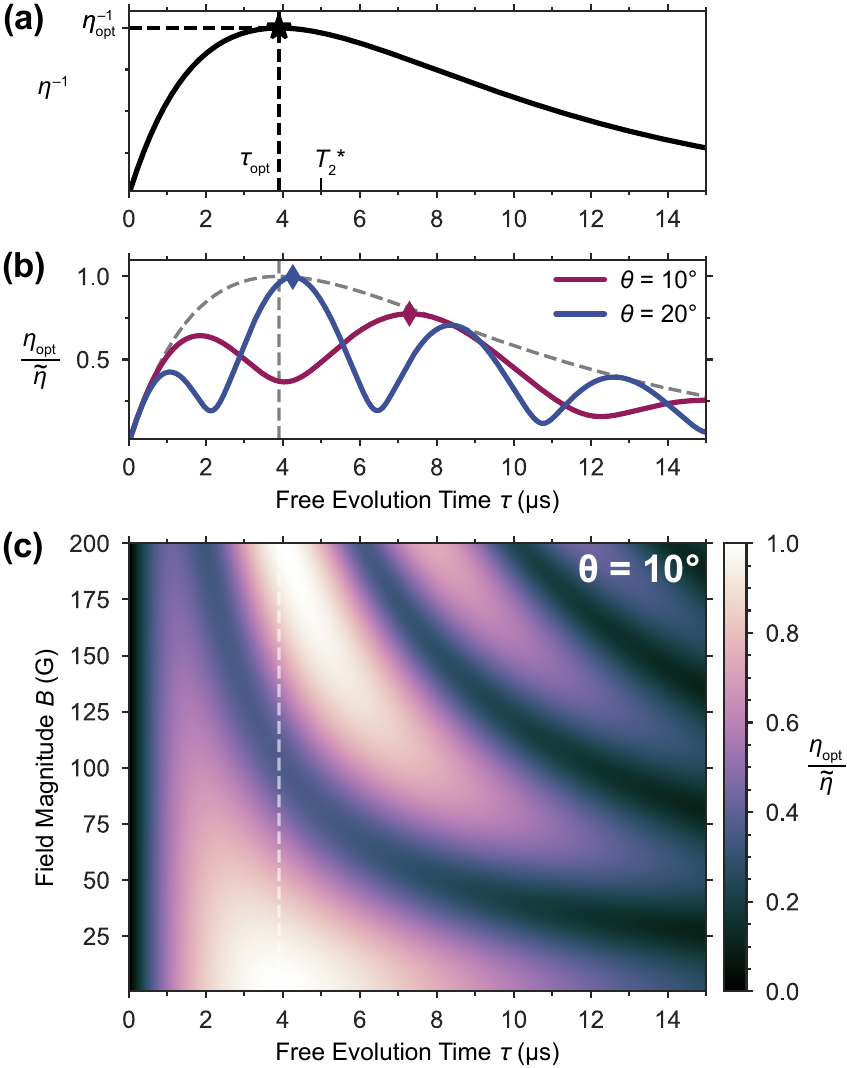}
    \caption{Calculations of expected Ramsey magnetic field sensitivity with and without envelope modulation (EREEM), and at different bias magnetic field configurations. (a) Inverse sensitivity $\eta^{-1}$ vs Ramsey free evolution time $\tau$ without EREEM, given by Eq.~\eqref{eq:etarelative} for fixed dead time $T_D=5\,$\textmu s and dephasing time $T_2^*=5\,$\textmu s. The optimal inverse sensitivity $\eta_\text{opt}^{-1}$ is indicated, corresponding to a free evolution time $\tau_\text{opt}$ indicated in all subfigures with a vertical dashed line. (b) Relative inverse sensitivity with EREEM $\eta_\text{opt}/\widetilde{\eta}$, normalized to the optimal sensitivity and free evolution time established in (a), for two distinct magnetic field misalignment angles $\theta=10^\circ$, $20^\circ$ at the fixed magnitude $B=100$\,G. Adjusted optimal evolution times $\widetilde{\tau}_\text{opt}$ are indicated by symbols $\Diamond$. (c) 2D color plot of the normalized inverse sensitivity at a fixed misalignment angle $\theta=10^\circ$, across bias field magnitudes $0<B<200$\,G.}
    \label{fig:4}
\end{figure}

To study the impact of EREEM on NV magnetic field sensitivity, we first consider a conventional Ramsey magnetometry measurement in the absence of any envelope modulation, using a single-quantum coherence between states $\ket{m_s=0}$ and either $\ket{m_s=+1}$ or $\ket{m_s=-1}$. A fixed Ramsey free evolution time $\tau$ is employed to map small magnetic field deviations onto changes in the fluorescence contrast. This working point is determined by optimizing the photon shot noise-limited magnetic field sensitivity \cite{Barry2020},
\begin{equation}
\label{eq:etashotnoise}
%\frac{1}{\Delta m}
\eta = \frac{1}{\gamma_e}\frac{1}{Ce^{-(\tau/T_2^*)^p}\sqrt{N}}\frac{\sqrt{\tau+T_{D}}}{\tau}.
\end{equation}
The average photon number is denoted by $N$ and the dead time $T_{D}$ represents the time spent outside the Ramsey sensing sequence for spin state initialization and readout. The maximum contrast $C$ decays due to NV spin dephasing, via a correction factor
$e^{-(\tau/T_2^*)^p}$. 
A plot of $\eta^{-1}$ as a function of $\tau$ is shown in Figure~\ref{fig:4}(a), setting $p=1$, $T_2^*=5\,$\textmu s, and $T_D=5\,$\textmu s according to our experimental conditions. This reveals
an optimal sensitivity $\eta_\text{opt}$ obtained at a corresponding free evolution time
$\tau_\text{opt}$, which approaches $T_2^*$ in the limit of long overhead time $T_D \gg \tau$. 

If EREEM is observed, then the contrast $C$ takes on an additional correction factor due to the amplitude modulation $\chi(\tau)$, which results in an adjusted sensitivity $\widetilde{\eta}$. Normalizing to $\eta_\text{opt}$, the relative inverse sensitivity is therefore given by the following ratio, assuming $N, T_2^*, p$, and $T_D$ remain unchanged between experiments:
\begin{equation}
\label{eq:etarelative}
\frac{\eta_\text{opt}}{\widetilde{\eta}} = 
\chi(\tau) \sqrt{\frac{\tau_\text{opt} + T_D}{\tau + T_D}} \exp\left(\frac{\tau_\text{opt}-\tau}{T_2^*}\right).
%\exp\Bigg(\frac{\tau_\text{opt}-\tau}{T_2^*} \Bigg)
\end{equation}
As established earlier, $\chi(\tau)$ oscillates between values $\chi_\text{min}=|\cos{\left(\Phi_{0,\pm 1}\right)}|$ and $\chi_\text{max}=1$, at a characteristic beat frequency $\omega_0$. These envelope properties are determined by the magnetic field magnitude $B$ and misalignment angle $\theta$, which in turn affect the relative inverse sensitivity $\eta_\text{opt}/\widetilde{\eta}$. 

The optimal sensitivity $\widetilde{\eta}_\text{opt} = \eta_\text{opt}$ is only achieved when an envelope maximum $\chi_\text{max}$ occurs at $\tau_\text{opt}$, obtained when the beat frequency $\omega_0$ is an integer multiple of $2\pi/\tau_\text{opt}$. If this condition is not satisfied, an updated optimal evolution time $\widetilde{\tau}_\text{opt}$
is necessary to minimize sensitivity degradation. These two scenarios are depicted in Fig.~\ref{fig:4}(b), which shows $\eta_\text{opt}/\widetilde{\eta}$ at two misalignment angles 10$^\circ$ and 20$^\circ$, for a field magnitude $B=100$\,G. The adjusted optimal evolution time $\widetilde{\tau}_\text{opt}$
for each case is marked, with a notable reduction in sensitivity seen in the 10$^\circ$ configuration. Similarly, changes in the field magnitude $B$ can affect sensitivity. Fixing the misalignment angle at 10$^\circ$ as an example (see the Supplemental Material for other misalignment angles \cite{SI}), Fig.~\ref{fig:4}(c) shows $\eta_\text{opt}/\widetilde{\eta}$ across a range of magnetic field magnitudes $B$ and free evolution times $\tau$.
Since the beat frequency $\omega_0$ increases with the field magnitude $B$, the relative inverse sensitivity $\eta_\text{opt}/\widetilde{\eta}$ exhibits faster oscillations with respect to $\tau$ for higher fields. The sensitivity is optimized at fields where a Ramsey envelope maximum coincides with $\tau_\text{opt}$, with the latter indicated by a dashed line in Fig.~\ref{fig:4}(c).

These calculations indicate that the sensitivity loss due to EREEM is highly dependent on changes in the bias field configuration. In practice, the tunability and control of such parameters depend on the specific sensing modality or application. For example, in experiments where an equal bias field projection on multiple NV axes is desired, the misalignment angle $\theta$ is highly constrained. Separately, there may be restrictions on the applied field magnitude $B$, for example, during studies of paramagnetic systems \cite{Glenn2017GGG}.

\section{Double-quantum Ramsey}

\begin{figure}[t]
    \centering
    \includegraphics[width=\linewidth]{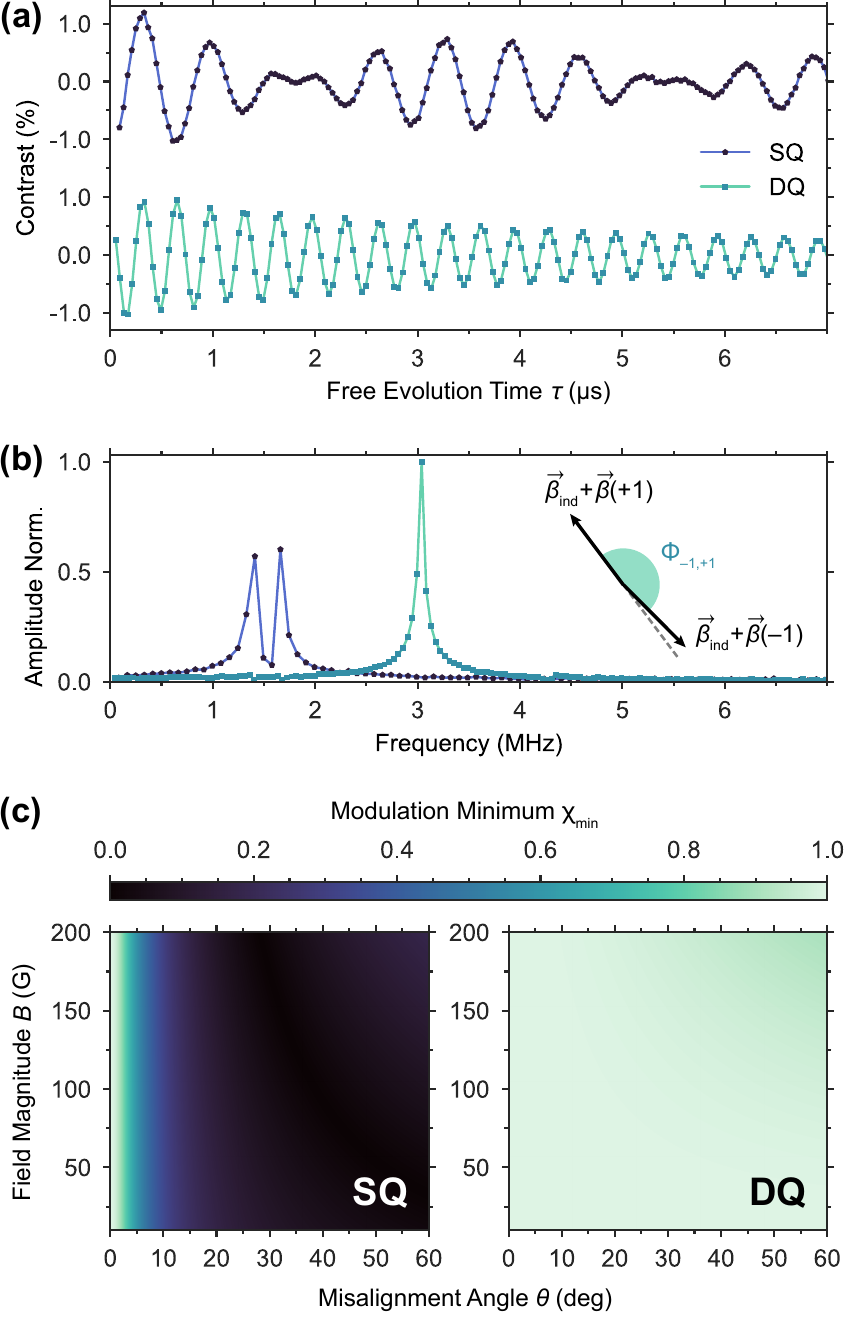}
    \caption{Comparison between experimental single-quantum (SQ) and double-quantum (DQ) Ramsey measurements under a misaligned bias magnetic field. Representative SQ and DQ Ramsey data collected at a bias field magnitude $B\approx50$\,G and misalignment angle $\theta\approx35^\circ$, in both time~(a) and frequency~(b) domains. The inset in (b) shows the effective nuclear magnetic fields $\vec{\beta}_\text{ind} + \vec{\beta}(\pm1)$ associated with the electronic spin states $\ket{m_s=\pm1}$ participating in the DQ protocol. The effective fields are nearly antiparallel ($\Phi_{-1,+1}\approx\pi$), %, with relative angle ,
    resulting in no observable envelope modulation (EREEM) in the DQ Ramsey signal ($\chi_\text{min}=|\cos\Phi_{-1,+1}|\approx 1$).
    %The DQ Ramsey signal does not exhibit observable envelope modulation, and accumulates a magnetic phase at twice the rate of the SQ oscillation.% due to the coherence between $m_s=-1$ and $+1$.
    (c) Calculation of relative contrast at amplitude modulation nodes $\chi_\text{min}$ for both SQ and DQ Ramsey signals, over a range of bias magnetic field magnitudes and misalignment angles. For SQ Ramsey, calculated values of $\chi_\text{min}=|\cos\Phi_{0,+1}|=|\cos\phi_{+1}|$ are shown, obtained using Eq.~\eqref{eq:phims}. For DQ Ramsey, $\chi_\text{min}=|\cos\Phi_{-1,+1}|=|\cos{\left(\phi_{+1}-\phi_{-1}\right)}|$.}
    \label{fig:5}
\end{figure}

As described in the previous section, envelope modulation (EREEM) in single-quantum (SQ) Ramsey experiments depends on the bias magnetic field configuration, and can result in significant magnetic field sensitivity loss. Alternatively, double-quantum (DQ) Ramsey protocols, which exploit the full NV spin-1 system, can circumvent the deleterious effects of EREEM. In fact, we observe a dramatic reduction of envelope modulation while using DQ coherence magnetometry. This behavior is illustrated in Fig.~\ref{fig:5}(a,b), which depicts measured SQ and DQ Ramsey free induction decay signals and their corresponding power spectra, at the same magnetic field configuration.

The DQ Ramsey protocol employs dual-tone microwave pulses with frequencies resonant with both the electronic transitions $m_s=0\leftrightarrow\!+1$ and $m_s=0\leftrightarrow\!-1$, often referred to as DQ pulses.
Besides this change to the applied pulses, the DQ Ramsey sequence mirrors the SQ protocol and consists of a pair of DQ pulses separated by a free evolution interval $\tau$.
The first DQ pulse prepares an equal superposition of the electronic spin states $\ket{+1}$ and $\ket{-1}$. After the interval $\tau$, a second DQ pulse maps the relative phase accumulated by these basis states onto the NV spin population, which is then read out optically. 

The lack of EREEM in the observed DQ signal can be understood by referring back to the vector model established in Sec. \ref{sec:theory}. The expected DQ Ramsey response can be described by Eq.~\eqref{eq:EREEMsig} after substituting the electronic basis states denoted by $i$ and $j$ with $-1$ and $+1$, respectively.
The effective nuclear magnetic fields associated with the electronic spin states $\ket{m_s=\pm1}$ are then given by $\vec{\beta}_{\text{ind}} + \vec{\beta}(\pm1)$ as depicted in Fig.~\ref{fig:5}(b). These field vectors are nearly antiparallel ($\Phi_{-1,+1}\approx\pi$), resulting in negligible envelope modulation $\chi_\text{min}=|\cos{\left(\Phi_{-1,+1}\right)}|\approx 1$ for a DQ Ramsey measurement.

The stark difference in envelope modulation behavior between SQ and DQ Ramsey is highlighted by calculations shown in Fig.~\ref{fig:5}(c). For a range of bias field magnitudes $B$ and misalignment angles $\theta$, the relative contrast at amplitude modulation nodes $\chi_\text{min}$ is plotted for both cases (see the Supplemental Material \cite{SI} for extended ranges of $B$ and $\theta$). Consistent with the results shown in Fig.~\ref{fig:3}(c),
the SQ Ramsey contrast at envelope nodes decays rapidly as a function of $\theta$, nearing zero even for small misalignment angles $\theta\sim10^\circ$. On the other hand, the DQ contrast is well preserved, with values of $\chi_\text{min}\approx 1$ across the fields considered in this work. 
Compared to SQ Ramsey, DQ Ramsey often provides more than an order of magnitude suppression of Ramsey amplitude modulation.

\section{Conclusion}
In this work, we present a physical model of electron Ramsey envelope modulation (EREEM) and find reasonable agreement with experimental measurements using an ensemble of $^{15}$NV centers in diamond. The observed envelope modulation exhibits a characteristic beat frequency and amplitude, dependent on the bias field magnitude and angle with respect to the NV quantization axis. We note a small systematic discrepancy between measurements of the envelope beat frequency and analytical predictions, which can be reconciled using an adjustment to the transverse hyperfine parameter $A_\perp$. These estimates deviate by around 3\%
%nearly 10\% 
from previous EPR measurements conducted at $\sim 2000$\,G \cite{Felton2009}, an order of magnitude greater than the magnetic fields considered here. However, these estimates of $A_\perp$ exhibit a dependence on the applied field magnitude, warranting additional measurements across an extended range of magnetic fields in future studies.

For magnetic field sensing modalities requiring misaligned magnetic fields, the integration of $^{15}$NV diamond and Ramsey coherence magnetometry is hindered by envelope modulation effects. We show that the resultant loss in sensitivity can be recovered by careful choice of the bias field. However, experimental constraints can limit the tunability of these field parameters. Alternatively, we find that double-quantum coherence magnetometry dramatically suppresses envelope modulation, while providing robustness to strain and temperature changes. 

\begin{acknowledgements}
This work was supported by the Army Research Laboratory MAQP program under Contract No. W911NF-19-2-0181, the DARPA DRINQS program under Grant No. D18AC00033, and the University of Maryland Quantum Technology Center. K.S.O. acknowledges support through an appointment to the Intelligence Community Postdoctoral Research Fellowship Program at the University of Maryland, administered by Oak Ridge Institute for Science and Education through an interagency agreement between the U.S. Department of Energy and the Office of the Director of National Intelligence.
%J. T. O. and J. T. contributed equally to this work.
\end{acknowledgements}

\bibliography{N15mag.bib}

\newpage
\onecolumngrid

\section*{Supplementary Material: Ramsey Envelope Modulation in NV Diamond Magnetometry}
%\appendix
\setcounter{section}{0}
\renewcommand{\thesection}{S-\Roman{section}}
\renewcommand{\theequation}{S.\arabic{equation}}
\renewcommand{\thefigure}{S\arabic{figure}}

\section{Theoretical Details} 
\label{app:theory}
This section provides a detailed derivation of the expressions given in Sec. \ref{sec:theory} of the main text. We start with the NV ground state Hamiltonian $H$, including the effects of a magnetic field restricted to the $x$-$z$ plane and the hyperfine interaction between the NV electronic spin and the onsite nitrogen nuclear spin,
\begin{equation}
\begin{aligned}
\frac{H}{\hbar} = D S_z^2 - \gamma_e \left( B_z \hat{S}_z +  B_x \hat{S}_x\right)  - \gamma_n g_n \left(B_z \hat{I}_z + B_x \hat{I}_x \right) +  A_{||} \hat{S}_z\hat{I}_z + A_{\perp} \left(\hat{S}_x\hat{I}_x+\hat{S}_y\hat{I}_y\right).
\end{aligned}
\end{equation}
The contributions to $H_0$ can be separated into secular and non-secular contributions $H_S$ and $H_{NS}$, respectively. The secular term $H_S$ consists of contributions that commute with operator $\hat{S}_z$. Conversely, the non-secular terms $H_{NS}$ do not commute with $\hat{S}_z$ and are capable of driving transitions within the electronic spin system. This is summarized as follows: 
\begin{equation}
\begin{aligned}
H &= H_S + H_{NS} \\
\frac{H_S}{\hbar}  &\equiv D\hat{S}_z^2 - \gamma_e   B_z \hat{S}_z - \gamma_n g_n \left(B_z \hat{I}_z +  B_x \hat{I}_x \right) + A_{||} \hat{S}_z\hat{I}_z\\
\frac{H_{NS}}{\hbar}  &\equiv -  \gamma_e   B_x \hat{S}_x + A_{\perp}\left(\hat{S}_x\hat{I}_x+\hat{S}_y\hat{I}_y\right).
\end{aligned}
\end{equation}

We begin by reproducing work done by 
Childress \textit{et al.} \cite{Childress2006} and Myers \cite{myers2016quantum}. The Hamiltonian is expanded as a series determined by order $D$, the largest energy scale within the Hamiltonian for the magnetic fields $B<200$\,G used in our study. The non-secular terms $H_{NS}$ are treated as a perturbation to the secular Hamiltonian $H_{S}$. Using second order perturbation theory, we obtain the leading order correction to the Hamiltonian  $H^{(1)}_{m_s}$ for a given electron spin quantum number $m_s$,
\begin{equation}
\begin{aligned}
\hat{P}_{m_s}H_{S} + H^{(1)}_{m_s}  = \hat{P}_{m_s} H_{NS} \frac{1}{E_{m_s} - \left(\mathbb{1}- \hat{P}_{m_s} \right)H_S\left(\mathbb{1}- \hat{P}_{m_s} \right)}H_{NS}\hat{P}_{m_s},
\end{aligned}
\end{equation}
with projection operator $\hat{P}_{m_s} \equiv  \ket{m_s}\bra{m_s}$ and identity operator $\mathbb{1}$.
An effective nuclear g-tensor $g_n(m_s)$ is defined by evaluating the change in the Hamiltonian with respect to the magnetic field,
\begin{equation}
\begin{aligned}
g_n(m_s) &= \frac{d}{dB}\left(\hat{P}_{m_s}H_{S} + H^{(1)}_{m_s}\right)
\\  &= 
\begin{pmatrix}
1 & 0 & 0\\
0 & 1 & 0\\
0 & 0 & 1
\end{pmatrix}
- \frac{\gamma_e}{\gamma_n D}(2 - 3 |m_s|)
\begin{pmatrix}
A_\perp & 0 & 0\\
0 & A_\perp & 0\\
0 & 0 & 0
\end{pmatrix}.
\end{aligned}
\end{equation}
This effective nuclear g-tensor is substituted into the expression for $H_S$. After introducing an enhancement parameter $\kappa=\frac{\gamma_e A_\perp}{\gamma_n D}\approx 8.26$, the following Hamiltonian is obtained:
\begin{equation}
\begin{aligned}
H_1 = &D\hat{S}_z^2 - \gamma_e B_z \hat{S}_z + A_{||} \hat{S}_z \hat{I}_z - 3\kappa \gamma_n B_x \hat{S}_z^2 \hat{I}_x \\&- \gamma_n \left(B_z \hat{I}_z + (1-2\kappa)  B_x \hat{I}_x\right).
\label{eq:Hleading}
\end{aligned}
\end{equation}
%After moving into an electronic spin double-rotating frame using rotation $R_1(t) \equiv exp\left(i 
%\big(\begin{smallmatrix}
%  \omega_e^{+} & 0 & 0\\
%  0 & 0 & 0\\
%  0 & 0 & \omega_e^{-}
%\end{smallmatrix}\big)
% t\right)$ with $\omega_e^{\pm} = D \pm \gamma_e B_z$, we obtain the Hamiltonian $\widetilde{H}_{1}$ given in Eq.~\eqref{eq:erotframe}. Experimentally this Hamiltonian is realized using microwave fields to probe transitions $m_s=0 \leftrightarrow +1$ and $m_s=0 \leftrightarrow -1$, both equally detuned from the two hyperfine resonances
%After moving into an electronic rotating frame using the unitary transformation $R_1(t) \equiv e^{i \omega_e \hat{S}_z t}$, with either $\omega_e = D \pm \gamma_e B_z$, we obtain the Hamiltonian $\widetilde{H}_{1}$ given in Eq.~\eqref{eq:erotframe}. 
After moving into an electronic doubly-rotating frame \cite{Mamin2014} using the unitary transformation 
\begin{equation}
\begin{aligned}
R_1(t) &\equiv \Big(e^{i \lambda_1 t}\ket{m_s=+1}\bra{m_s=+1} + \ket{m_s=0}\bra{m_s=0} + e^{i \lambda_2 t}\ket{m_s=-1}\bra{m_s=-1} \Big) \otimes \mathbb{1}_n \\
&= \begin{pmatrix}
e^{i \lambda_1 t} & 0 & 0\\
0 & 1 & 0\\
0 & 0 & e^{i \lambda_2 t}
\end{pmatrix} \otimes 
\begin{pmatrix}
1 & 0 \\
0 & 1
\end{pmatrix}
\end{aligned},
\end{equation}
with $\lambda_{1,2} = D \mp \gamma_e B_z$, we obtain the Hamiltonian $\widetilde{H}_{1}$ given in Eq.~\eqref{eq:erotframe}, which is also given in the main text as Eq.~\eqref{eq:Hrot}.
Experimentally this Hamiltonian is realized using microwave fields to probe  transitions $\ket{m_s=0} \leftrightarrow \ket{m_s=+1}$ and/or $\ket{m_s=0} \leftrightarrow \ket{m_s=-1}$, equally detuned from the two hyperfine resonances.
\begin{equation}
\begin{aligned}
\widetilde{H}_{1} &= 
R_1(t) H_{1} R_1(t)^\dagger + i \frac{d R_1(t)}{dt} R_1(t)^\dagger
\\&= \underbrace{A_{||} \hat{S}_z \hat{I}_z - 3\kappa \gamma_n B_x \hat{S}_z^2 \hat{I}_x}_\text{depends on $m_s$}- \underbrace{\gamma_n \left(B_z \hat{I}_z + (1-2\kappa)  B_x \hat{I}_x\right)}_\text{independent of $m_s$}
\label{eq:erotframe}
\end{aligned}
\end{equation}
An additional rotation using $R_2(t) = e^{i \phi \hat{I}_y}$ further simplifies the Hamiltonian, with $\phi = \tan^{-1}{\left( \frac{(1-2\kappa) B_x}{B_z}\right)}$. This rotation diagonalizes the nuclear subsystem that is independent of the electron spin, producing the following Hamiltonian:
\begin{equation}
\begin{aligned}
\widetilde{H}_{2} &= 
R_2(t) \widetilde{H}_{1} R_2(t)^\dagger + i \frac{d R_2(t)}{dt} R_2(t)^\dagger \\
&= -\gamma_n %\underbrace{
\left( \vec{\beta}_{\text{ind}} + \vec{\beta}(m_s) \right)
\cdot \vec{I}.
\label{eq:diagnuc}
\end{aligned}
\end{equation}
Equation \eqref{eq:diagnuc}, also provided in the main text as Eq.~\eqref{eq:Hnuc}, consists of an effective magnetic field $\vec{\beta}_{\text{ind}} + \vec{\beta}(m_s)$ interacting with the nuclear spin. 
This vector sum consists of: 
\begin{enumerate}
\item a magnetic field along the newly defined $z'$-axis, \textit{independent} of the electronic spin state,
\begin{equation}
\label{eq:B0}
\begin{aligned}
\vec{\beta}_{\text{ind}} = \beta_{\text{ind}} \hat{z'}= \sqrt{B_z^2 + (1-2\kappa)^2 B_x^2}\,\hat{z'},
\end{aligned}
\end{equation}
\item and a spin-\textit{dependent} field
\begin{equation}
\label{eq:Bms}
\begin{aligned}
\vec{\beta}(m_s) =
\begin{bmatrix}
\beta_{x'}(m_s)\\
\beta_{y'}(m_s)\\
\beta_{z'}(m_s)
\end{bmatrix} = 
\frac{1}{\beta_{\text{ind}}}
\begin{bmatrix}
B_x \left( m_s(1-2\kappa)\frac{ A_{||} }{\gamma_n} + m_s^2 3 \kappa B_z \right)
\\
0\\
-m_s\frac{A_{||}}{\gamma_n}B_z + m_s^2 3 \kappa (1-2 \kappa) B_x^2
\end{bmatrix}.
\end{aligned}
\end{equation}
\end{enumerate}
An additional rotation generated by $\hat{I}_{y'}$ by an angle $\phi_{m_s}$ diagonalizes the Hamiltonian, depending on the electronic spin state $m_s$,
\begin{equation}
\begin{aligned}
\label{eq:phims_supp}
\phi_{m_s} = \tan^{-1}{\left(\frac{\beta_{x'}(m_s)}{\beta_{\text{ind}} + \beta_{z'}(m_s)}\right)}.
\end{aligned}
\end{equation}
From Eq.~\eqref{eq:diagnuc}, the effective nuclear Larmor precession frequency can be obtained,
\begin{equation}
\begin{aligned}
\omega_{m_s} = \left\vert\gamma_n  \left(\vec{\beta}_{\text{ind}} + \vec{\beta}(m_s) \right)\right\vert &= -\gamma_n \sqrt{\Big(\beta_{\text{ind}} + \beta_{z'}(m_s)\Big)^2 + \beta_{x'}(m_s)^2}\\
&= -\gamma_n \sqrt{B_x^2\Big(1 - 2\kappa+m_s^2 3  \kappa\Big)^2 + \left(m_s \frac{A_{||}}{\gamma_n} - B_z\right)^2}.
\end{aligned}
\end{equation}
For a given electronic spin state $m_s$, the nuclear spin precesses at a corresponding frequency $\omega_{m_s}$. Note that when $m_s=0$, this reduces to the spin-independent Larmor frequency $\omega_0 = \left|\gamma_n \beta_{\text{ind}} \right|$.

For a single-quantum (SQ) Ramsey sequence, we effectively operate within a 2-level system consisting of $\ket{m_s=0}$ and either $\ket{m_s=+1}$ or $\ket{m_s=-1}$. To apply a microwave rotation pulse, we define the unitary operator $\hat{U}_x(\theta) = e^{-i \hat{\sigma}_x \theta/2}$ using the 2-level Pauli spin operator $\hat{\sigma}_x$. The SQ Ramsey sequence consists of two $\pi/2$ pulses separated by a free precession period of duration $\tau$. The initial state $\ket{\psi(0)}=\ket{m_s=0}$
evolves under the following unitary rotations:
\begin{equation}
\begin{aligned}
\psi_{SQ}(t) = \hat{U}_x\left(\pi/2\right)
\left(e^{-i \widetilde{H}_{2} \tau/\hbar}\right)
\hat{U}_x\left(\pi/2\right)\ket{\psi(0)}.
\label{eq:Rpulseseq}
\end{aligned}
\end{equation}
%\exp{\left(\frac{-i \widetilde{H}_{2} \tau}{\hbar}\right)}
The Ramsey response $S_{SQ}$ can be represented by a projection $P_0=\ket{m_s=0}\bra{m_s=0}$ onto the $m_s=0$ state,
\begin{equation}
\begin{aligned}
\label{eq:R,SQsimp}
S_{SQ}(\tau) &= \text{Tr}\left\{P_0 \ket{\psi_{SQ}(\tau)}\bra{\psi_{SQ}(\tau)}\right\}
\\ &= 
\frac{1}{2} \bigg(
1-\cos{\left(\frac{\omega_0 \tau}{2}\right)} \cos{\left(\frac{\omega_{\pm1} \tau}{2}\right)}
+
\cos{\left(\Phi_{0, \pm 1}\right)}
 \sin{\left(\frac{\omega_0 \tau}{2}\right)} \sin{\left(\frac{\omega_{\pm1} \tau}{2}\right)}
\bigg).
\end{aligned}
\end{equation} 
This response represents the electronic spin population and therefore spans between $0$ and $1$. For tidiness, this form is slightly modified to give the expression in the main text $S_{i,j}(\tau) = 2 S(\tau)-1$, where $S(\tau)$ denotes the response shown in Eq.~\eqref{eq:R,SQsimp}. The double-quantum (DQ) response can be similarly obtained by substituting the 2-level spin operator $\sigma_x$ in the unitary rotation $\hat{U}(\theta)$ with the appropriate spin-1 operator:
\begin{equation}
\label{eq:R,DQsimp}
S_{DQ} = 
\frac{1}{2}\left[
1+\cos{\left(\frac{\omega_{-1} \tau}{2}\right)} \cos{\left(\frac{\omega_{+1} \tau}{2}\right)}
-\cos{\left(\Phi_{-1,+1}\right)} \sin{\left(\frac{\omega_{-1} \tau}{2}\right)} \sin{\left(\frac{\omega_{+1} \tau}{2}\right)}
\right]
\end{equation}

\section{Experimental Details}
\label{app:exp}
A  diode‐pumped solid‐state laser (Lighthouse Photonics, Sprout-D5W) is employed to generate the continuous-wave 532\,nm laser beam. The laser beam is passed through an acousto-optic modulator (AOM) (Gooch \& Housego, Model: 3250-220) and the first order diffracted beam is used for the experiments. To create optical pulses for NV ensemble initialization and readout, the RF driver for the AOM is gated by switches (Mini-Circuits, ZASWA-2-50DR+) using transistor-transistor logic (TTL) pulses (Swabian Instruments, Pulse Streamer 8/2).
The beam is focused through a 20$\times$/0.75\,NA Nikon objective onto the diamond, illuminating the NV-enriched layer over a spot of $\sim20\,$\textmu m diameter. The illumination intensity is kept below 0.1\,mW/\textmu m$^{2}$. The outgoing spin-dependent NV fluorescence is collected through the same objective, passed through a $150\,$\textmu m pinhole (Thorlabs, P150D) to restrict the collection volume, filtered by a 647\,nm long pass filter (Semrock) and finally projected onto an avalanche photodiode (Hamamatsu C10508-01). Photodiode voltage measurements are acquired using a DAQ system (National Instruments, NI USB-6363). The bias magnetic field is generated using two permanent samarium cobalt (SmCo) ring magnets to minimize temperature sensitivity of the bias field. Rotation of the magnet pair for different misalignment angles is controlled using two sets of Thorlabs HDR50 rotation stages with stepper motors. Microwave fields are synthesized from up to two signal generators with built-in IQ mixers (Stanford Research System, SG384); and gated by switches and TTLs to provide either single- or double- quantum pulsed control of the NV ensemble spin states. The $\Omega$-shaped planar waveguide for microwave delivery has a diameter $\approx150\,$\textmu m to ensure field homogeneity over the laser illumination area. 

\section{Bias Field Misalignment Angle} \label{app:Biasfieldmisalignmentangle}

\begin{figure}[h!]
    \centering
    \includegraphics[width=8.6 cm]{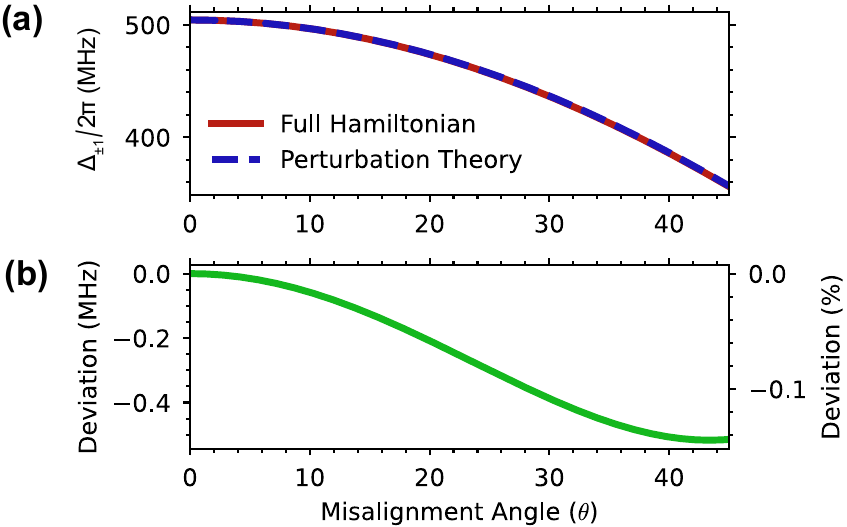}
    \caption{Transition frequency difference $\Delta_{\pm1}$ between $m_s=0\leftrightarrow+1$ and $m_s=0\leftrightarrow-1$ resonances as a function of misalignment angle. Calculation is performed at 90\,G. (a) Numerical results of $\Delta_{\pm1}$ using the full NV ground state Hamiltonian (solid, red) and the approximate formula from perturbation theory $2\gamma_e B \cos{\theta}$ (dashed, blue). (b) Absolute and percentage deviation of $\Delta_{\pm1}$ between the two calculation methods.
    }\label{fig:angleest}
    %\label{fig: A2}
\end{figure}

When the bias magnetic field is misaligned from the NV axis, both the transverse $B_x$ and longitudinal $B_z$ field components affect the energy levels of $\ket{0}, \ket{\pm1}$. In the main text, the resonance frequency difference $\Delta_{\pm1}$ between the two transitions $\ket{m_s=0}\leftrightarrow\ket{m_s=+1}$ and $\ket{m_s=0}\leftrightarrow\ket{m_s=-1}$ is used to calculate the misalignment angle $\theta$, via the relationship $\Delta_{\pm1} \approx 2\gamma_e B_{z}=2\gamma_e B \cos{\theta}$. This approximate relationship is obtained from time-independent perturbation theory. For the small bias field magnitudes $B\lesssim110$\,G used in our study, the large NV zero field splitting $D=2\pi\times2870$\,MHz dominates the energy scale, $D\gg\left\vert\gamma_e B\right\vert$. This allows us to treat the transverse field $B_x$ contribution as a perturbation, resulting in the eigenenergies
$E_{\ket{0}\rightarrow\ket{\pm1}}$ accurate to the second order correction \cite{Bauch2018}:
\begin{equation}
\label{eq:updatedTransitionFrequency}
\frac{E_{\ket{0}\rightarrow\ket{\pm1}}}{\hbar} =D + \frac{3\left|\gamma_e B_x\right|^2}{D} \pm \gamma_e B_z.
\end{equation}
Therefore, $\Delta_{\pm1}=\left(E_{\ket{0}\rightarrow\ket{+1}}-E_{\ket{0}\rightarrow\ket{-1}}\right)/\hbar$ can be accurately modeled by $2\gamma_e B \cos{\theta}$, up to second order. Using this relationship, the measurements of $\Delta_{\pm 1}$ are converted to values of the misalignment angle $\theta$. The error bars for $\theta$ in Fig.~\ref{fig:3} of the main text indicate the corresponding 95\% confidence interval (CI), except for cases where the standard deviation $\sigma_\theta$ calculated from error propagation is less than the stage rotation error. In these cases, the nominal accuracy ($\pm 0.047^\circ$) of the stage is used for $\sigma_\theta$ instead.

To verify that the expression for $\Delta_{\pm1}$ is indeed valid, we compare the measured values of $\Delta_{\pm1}$ to those obtained by diagonalizing the complete Hamiltonian given in Eq.~\eqref{eq:HGSlab}. An example plot is shown in Fig.~\ref{fig:angleest}, for a 90\,G bias field magnitude and misalignment angles $\theta$ up to 45$^\circ$. We observe excellent agreement with the results obtained using the full Hamiltonian, with percentage deviation in $\Delta_{\pm1}$ of under 0.15\%. 
The corresponding error contribution to the angle estimation is less than the dominant error values used for Fig. \ref{fig:3} in the main text.

\section{Microwave Frequency Calibration}
\label{app:mwcalib}

To set the microwave frequency to the center of the two $^{15}$NV hyperfine-split transitions, we perform a two-step calibration procedure combining pulsed-ODMR and Ramsey techniques. First, the two hyperfine transitions are resolved
by performing a pulsed-ODMR experiment with low microwave power ($\pi$ pulse time $\sim 900$\,ns). The resulting spectrum is fit to a sum of two Lorentzians $A-\sum_{i=1}^2\frac{C_i}{\pi}\frac{\frac{1}{2}\Gamma_i}{(\nu-\nu_i)^2+(\frac{1}{2}\Gamma_i)^2}$, with free parameters including the contrast $C_i$, line width $\Gamma_i$, frequency $\nu_i$ and overall offset $A$. 

The mean of the two Lorentzian center frequencies $\nu_* = (\nu_1 + \nu_2)/2$ is used as the initial value of the applied microwave frequency. Next, a series of Ramsey $\pi/2-\tau-\pi/2$ experiments are performed to refine this estimate.
Fixing $\tau$ at a point of maximum contrast, the microwave frequency is varied around $\nu_*$ to produce an oscillating magnetometry curve \cite{Barry2020}. Choosing a common phase for both microwave $\pi/2$ pulses, the extremum nearest to $\nu_*$ is used as the updated center resonance frequency. This calibrated value $\nu_*^{calib}$ is obtained by fitting the magnetometry curve to a sinusoidal function and extracting the necessary frequency offset.

\section{Measured Ramsey Time Series Fit Details}
\label{app:Ramseyfit}

%The modified form of used for fits to Ramsey time series data is

For fits to Ramsey time series data, we use a modified form of Eq.~\eqref{eq:EREEMsig} from the main text:
\begin{equation} \label{eq:completeereemfit}
\begin{aligned}
%&S(\tau, \omega_0, \omega_{+1}, \Phi_{0,+1}, T_2^*, p, x_0, x_{+1}, B, C_0)
%
&\widetilde{S}_{0,+1}(\tau)
\\&= 
C_0 e^{-(\tau/T_2^*)^p}\biggl[-\cos{\left(\frac{\omega_{0}}{2}\tau+x_0\right)}\cos{\left(\frac{\omega_{+1}}{2}\tau+x_{+1}\right)} +\cos{\left(\Phi_{0,+1}\right)}\sin{\left(\frac{\omega_{0}}{2}\tau+x_0\right)\sin{\left(\frac{\omega_{+1}}{2}\tau+x_{+1}\right)}} \biggr]+B.
\end{aligned}
\end{equation}
Here, $\tau$ represents the free evolution time in a Ramsey sequence. The free fit parameters are the initial contrast $C_0$, dephasing time $T_2^*$, stretched exponential factor $p$, EREEM frequency components $\omega_0$ and $\omega_{+1}$, amplitude modulation factor determined by $\Phi_{0,+1}$, phase offsets $x_{0}$ and $x_{+1}$, and a vertical offset $B$.

\begin{figure}[H]
    \centering
    \includegraphics[width=8.6 cm]{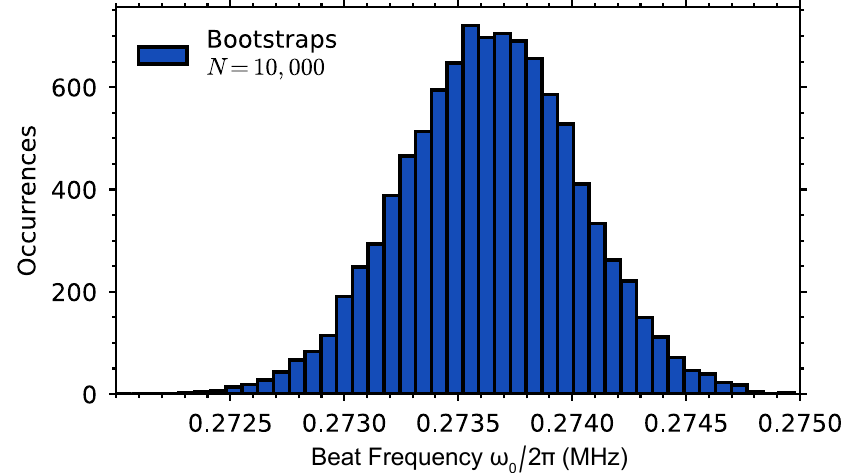}
    \caption{Example bootstrap fit result for the EREEM beat frequency. Starting with a single Ramsey time series, $10,000$ additional datasets are created by randomly sampling with replacement from the initial time series. Estimates of the EREEM beat frequency $\omega_0$ are extracted using the MATLAB \texttt{lsqcurvefit} function for each dataset. The distribution supports the assumption of normality used by the standard interval method for confidence interval (CI) construction.}
    \label{fig: bootstrap}
\end{figure}

For each estimate of $\omega_0$ or $\chi_\text{min}=\cos{\left(\Phi_{0,+1}\right)}$ reported in Fig.~\ref{fig:3}(b,c) of the main text, its corresponding confidence interval (CI) is constructed using the standard interval method. For each parameter estimate $\hat{\theta}$, its error bar represents a range bounded between $\hat{\theta}\pm z^{(0.95)} \hat{\sigma}$, where $\hat{\theta}$ represents the point estimate of the parameter of interest with standard error $\hat{\sigma}$, and $z^{(0.95)}$ denotes the 95$^{th}$ percentile of the normal deviate. Both $\hat{\theta}$ and $\hat{\sigma}$ are calculated using maximum likelihood estimation via the MATLAB \texttt{lsqcurvefit} function. However, the exact CI can differ from the standard interval approximation if the measurement error is not normally distributed. In such scenarios, a bootstrap confidence interval produces a more accurate CI estimation \cite{DiCiccio1996}. To test this approach, we applied a bootstrap sampling method to four randomly selected data points from Fig.~\ref{fig:3}(b,c) in the main text. The distribution of fitted parameters resembles a normal distribution with $\sim95$\% data points located within $\hat{\theta}\pm z^{(0.95)} \hat{\sigma}$, which supports the reporting of CI estimates obtained by the standard interval method in the main text. A bootstrap fitting example for the envelope beat frequency $\omega_0$ is shown in Fig.~\ref{fig: bootstrap}, for a bias field magnitude $B=90.08(2)\,$G and a misalignment angle $\theta=25.72(9)^\circ$.

\begin{figure}[H]
    \centering
    \includegraphics[width=14cm]{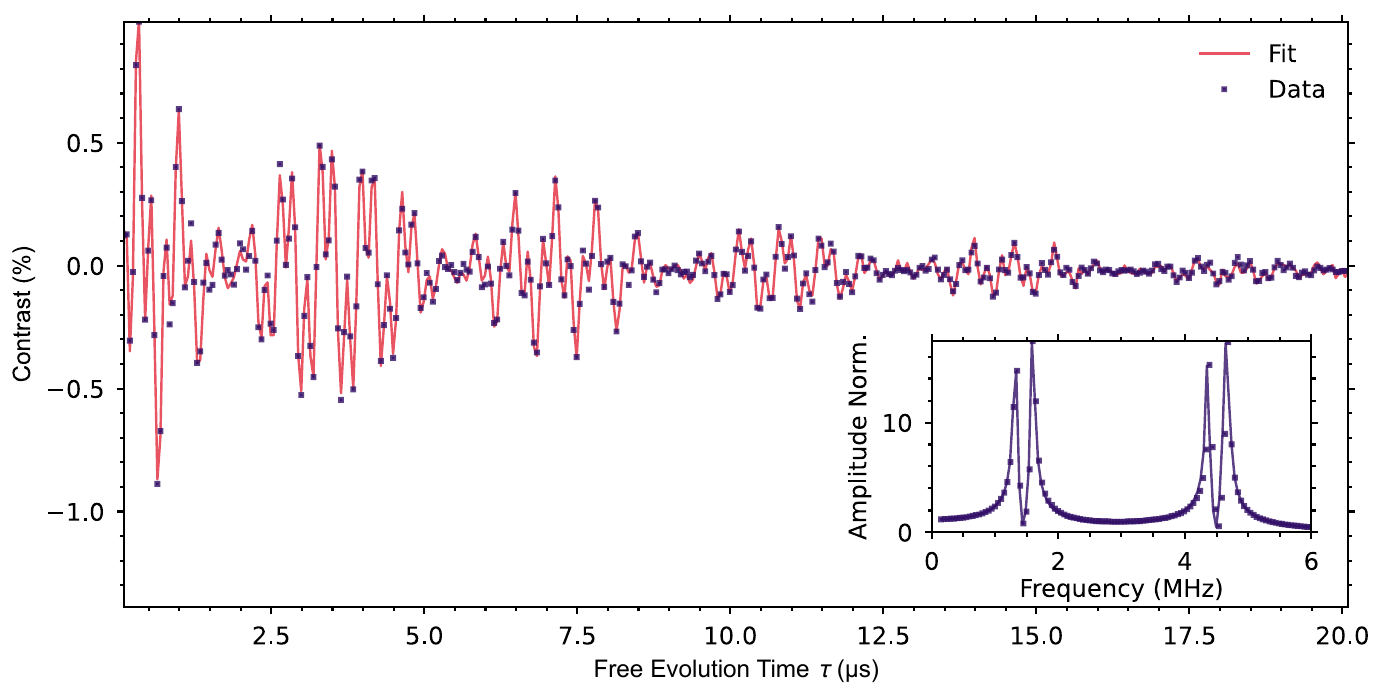}
    \caption{Ramsey oscillation with a purposefully detuned microwave driving frequency. The larger detuning enables clear separation among all frequencies components. The fit result for the EREEM beat frequency $\omega_0$ is $0.2733\pm0.0014$\,MHz, overlapping with the confidence interval (CI) of the corresponding beat frequency fit result, $0.2736\pm0.0008$\,MHz, using a microwave frequency equally detuned from the two hyperfine transitions.} \label{fig: fourfreqereem}
\end{figure}

A fit to Eq.~\eqref{eq:completeereemfit} is appropriate when the microwave driving frequency is equally detuned from the two $^{15}$NV hyperfine-split transitions, $|\delta_{m_I=-\frac{1}{2}}|=|\delta_{m_I=\frac{1}{2}}|$. Alternatively, the EREEM beat frequency $\omega_0$ can be extracted when these detunings are not equal $|\delta_{m_I=-\frac{1}{2}}|\neq|\delta_{m_I=\frac{1}{2}}|$. This approach comes at the expense of fitting accuracy due to the presence of extra frequency tones in the Ramsey response. In such a case, we expect a four-tone Ramsey oscillation with two frequency splittings of magnitude $\omega_0$ centered around $|\delta_{m_I=-\frac{1}{2}}|$ and $|\delta_{m_I=\frac{1}{2}}|$, respectively. To resolve all frequency components and to avoid ambiguity in the frequency assignments, the following condition is enforced: $|\delta_{m_I=-\frac{1}{2}}|-|\delta_{m_I=\frac{1}{2}}|\gg\omega_0$.
This alternative method is applied to four randomly selected field configurations from those shown in Fig.~\ref{fig:3} of the main text. A four-tone EREEM function modified from Eq.~\eqref{eq:completeereemfit} is used for fitting to the data. The extracted envelope beat frequencies overlap with the CI of the corresponding data under the same field configurations in the main text. Example data is shown in Fig.~\ref{fig: fourfreqereem}, for a bias field magnitude $B=90.08(2)$\,G and misalignment angle $\theta=25.72(9)^\circ$.

\section{QuTiP Simulations}
\label{app:qutipsimulation}

\begin{figure}[h!]
    \centering
    \includegraphics[width=17.8 cm]{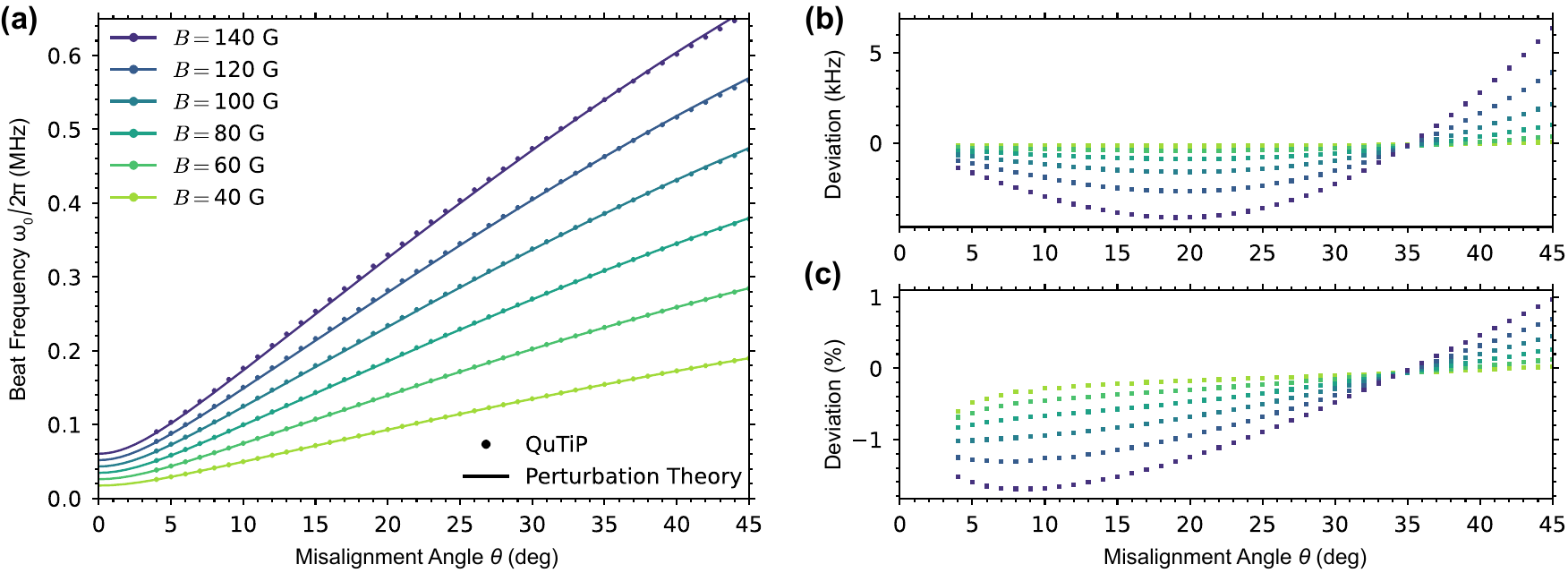}
    \caption{Calculation and comparison of Ramsey envelope (EREEM) beat frequency $\omega_0$ as a function of bias magnetic field magnitude and misalignment angle, using analytical approximation and QuTiP numerical simulations. (a) $\omega_0$ obtained using numerical simulations (circles) and analytical perturbation theory (solid line), for magnetic fields between $40$\,G and $140$\,G. (b) Absolute differences between $\omega_0$ estimates from QuTiP numerical simulations and analytical approximation, for the field configurations shown in (a). (c) Percentage deviation between $\omega_0$ estimates from QuTiP numerical simulations and analytical approximation.}
    \label{fig:qutipvstheory}
\end{figure}

As discussed in Sec. \ref{sec:envprops} of the main text, systematic differences are observed between experimental estimates of the Ramsey envelope (EREEM) beat frequency $\omega_0$, and theoretical predictions from Eq.~\eqref{eq:omega0} of the main text obtained using second order perturbation theory. This equation is reproduced below for clarity.
\begin{equation}
\tag{\ref{eq:omega0}}
\begin{aligned}
\omega_0
= \left|\gamma_n \beta_{\text{ind}}\right|
&=\left|\gamma_n\right|\sqrt{B_z^2 + (1-2\kappa)^2 B_x^2}\\
&=\left|\gamma_n\right| B\sqrt{1+ 4(\kappa^2-\kappa) \sin^2{\theta}}.
\end{aligned}
\end{equation}
The observed difference increases with the bias magnetic field misalignment angle. To study this discrepancy, we conduct numerical simulations of Ramsey spin dynamics for a range of bias field configurations, using the QuTiP package in Python.
For a coupled NV electron-nuclear system initially described by the lab frame Hamiltonian in Eq.~\eqref{eq:HGSlab} of the main text, applied microwave pulses are modeled using time-dependent contributions $\Omega(t) \cos{(\omega_e t + \varphi)}\hat{S}_x$. The AC magnetic field amplitude $\Omega(t)$ is chosen to yield a 20\,MHz Rabi frequency during pulses. The duration of each pulse is calibrated so a $\pi/2$ rotation is performed. The phase $\varphi$ is used to toggle between pulse rotation axes. 

For each field configuration, the lab frame Hamiltonian $H$ is first diagonalized to obtain eigenenergies and eigenstates. The pair of frequencies for transitions between $\ket{m_s=0,m_I +1/2}\leftrightarrow\ket{1,+1/2}$ and $\ket{0, -1/2}\leftrightarrow\ket{1,-1/2}$ are calculated. The mean of these two values is used as the frequency $\omega_e$ of the pulses.
Starting with an initial state $\ket{0, -1/2}$, the Ramsey pulse sequence is applied and the final population in the $m_s=0$ state is determined. Pulse sequences are simulated for a range of free evolution times $\tau$, and the resulting time series is fit to Eq.~\eqref{eq:R,SQsimp} to extract the parameters $\omega_0,$ $\omega_{\pm1}$ and $\Phi_{0,\pm1}$. In Fig.~\ref{fig:qutipvstheory}, the values of $\omega_0$ extracted from numerical simulations are compared to the analytical predictions given by Eq.~\eqref{eq:omega0}. For the field configurations considered in the main text, the difference between these calculations remains around or below $1\%$.

\section{Normalized Inverse Sensitivity}
\label{app:2Drelativecontrast}

\begin{figure}[h!]
    \centering
    \includegraphics[width=17.8 cm]{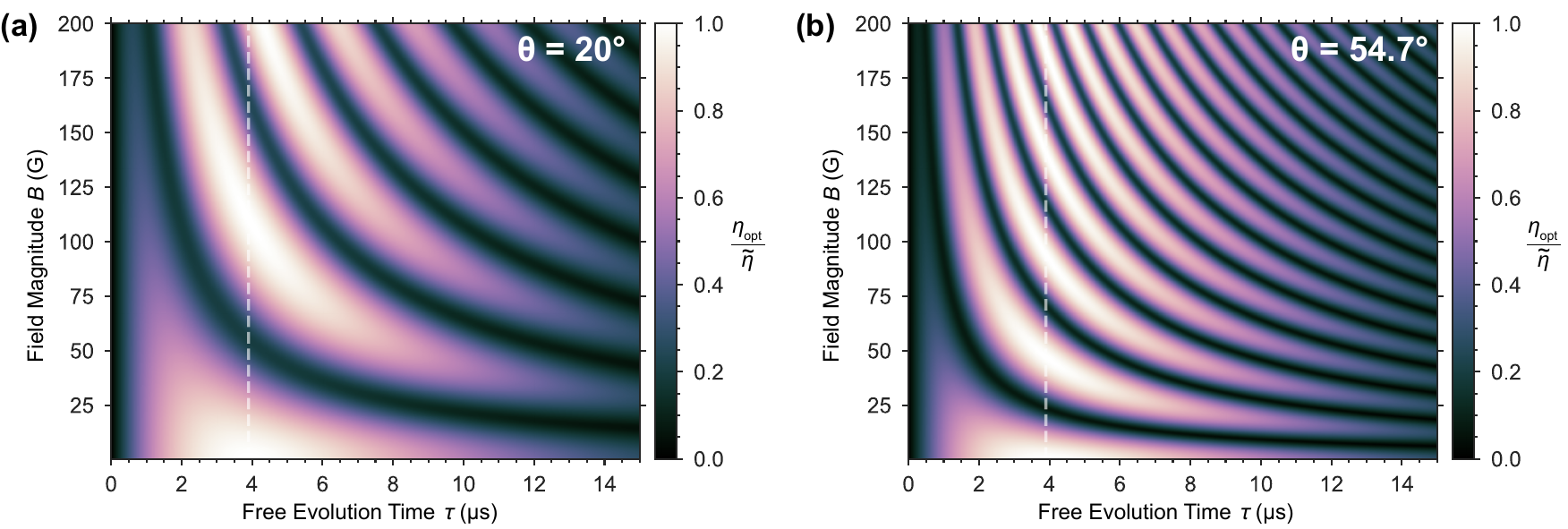}
    \caption{2D color plots of the normalized inverse sensitivity with respect to the Ramsey free evolution time $\tau$ and bias magnetic field magnitude $B$, at 2 distinct misalignment angles $\theta=20^\circ$ (a) and $54.7^\circ$ (b)} 
    \label{fig:2Drelativesensitivity}
\end{figure}

In Fig. \ref{fig:4}(c) of the main text, a 2D color plot of the relative normalized inverse magnetic field sensitivity is shown, evaluated using Eq. \eqref{eq:etarelative}. These calculations are repeated here for two other misalignment angles $\theta=20^\circ, 54.7^\circ$ and are shown in Fig. \ref{fig:2Drelativesensitivity}. The experimental timescales assumed here ($T_D=5\,$\textmu s, $T_2^*=5$\textmu s) are the same as those used in the main text. In particular, the $\theta=54.7^\circ$ case can describe a bias magnetic field oriented to have equal projections onto all four NV quantization axes.

\section{Transverse Hyperfine Parameter}
\label{app:transverhyperfine}

\begin{figure}[h!]
    \centering
    \includegraphics[width=8.6 cm]{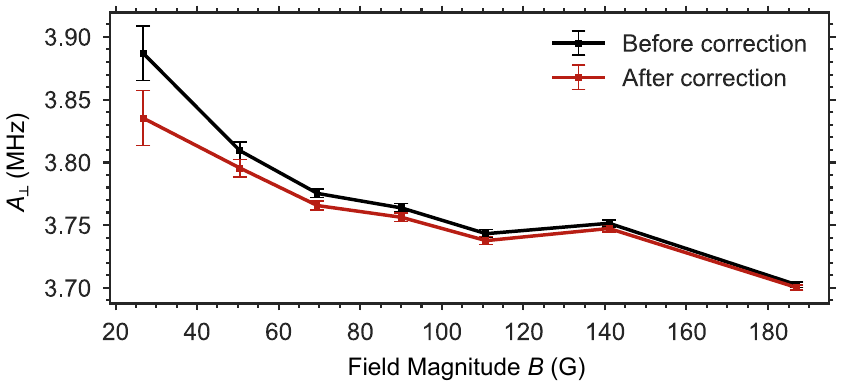}
    \caption{
    Results for the transverse hyperfine parameter $A_\perp$ as a fit parameter to the Ramsey envelope (EREEM) beat frequency $\omega_0$, extracted from experimental data, for a range of magnetic fields, with error bars showing the 95\% confidence interval, shown in black. Estimated corrections to values of $A_\perp$ are depicted in red, using the discrepancy between QuTiP numerical simulations and theoretical predictions of $\omega_0$.
    } 
    \label{fig:AperpvsB}
\end{figure}

As discussed in Sec. \ref{sec:envprops} of the main text, we perform fits of Eq.~\eqref{eq:omega0} to experimental measurements of $\omega_0$ while allowing the transverse hyperfine parameter $A_\perp$ to vary as a fit degree of freedom. We repeat this process at a series of bias field magnitudes, extending beyond those considered in the main text.
The resulting fits for $A_\perp$ are shown in Fig.~\ref{fig:AperpvsB} using black markers, with error bars representing a 95\% confidence interval (CI). We observe an overall decrease of $A_\perp$ as the field magnitude is increased. 

Due to the field-dependent discrepancy of $\omega_0$ between QuTiP simulations and the predictions of perturbation theory, additional corrections are necessary to precisely determine $A_\perp$. However, as shown in Fig.~\ref{fig:qutipvstheory}, the degree to which these predictions differ depends on the misalignment angle $\theta$.
As an initial attempt to estimate these corrections, we fit Eq.~\eqref{eq:omega0} to the estimates of $\omega_0$ obtained using QuTiP simulations, at the specific magnetic field magnitudes and misalignment angles studied in experiment. The absolute difference between these fits and the originally assumed value for $A_\perp$ is used to update the experimentally determined $A_\perp$ estimates, shown in Fig.~\ref{fig:AperpvsB} as red points.
%Therefore, 
We expect a future study to further refine the estimation for $A_{\perp}$ by increasing the sampling of misalignment angles at each magnetic field magnitude.
In addition, an extension of these studies to higher field magnitudes will allow direct comparisons to the previous ESR study \cite{Felton2009}.

\section{Double-quantum Envelope Amplitude}

\begin{figure}[h]
    \centering
    \includegraphics[width=8.6 cm]{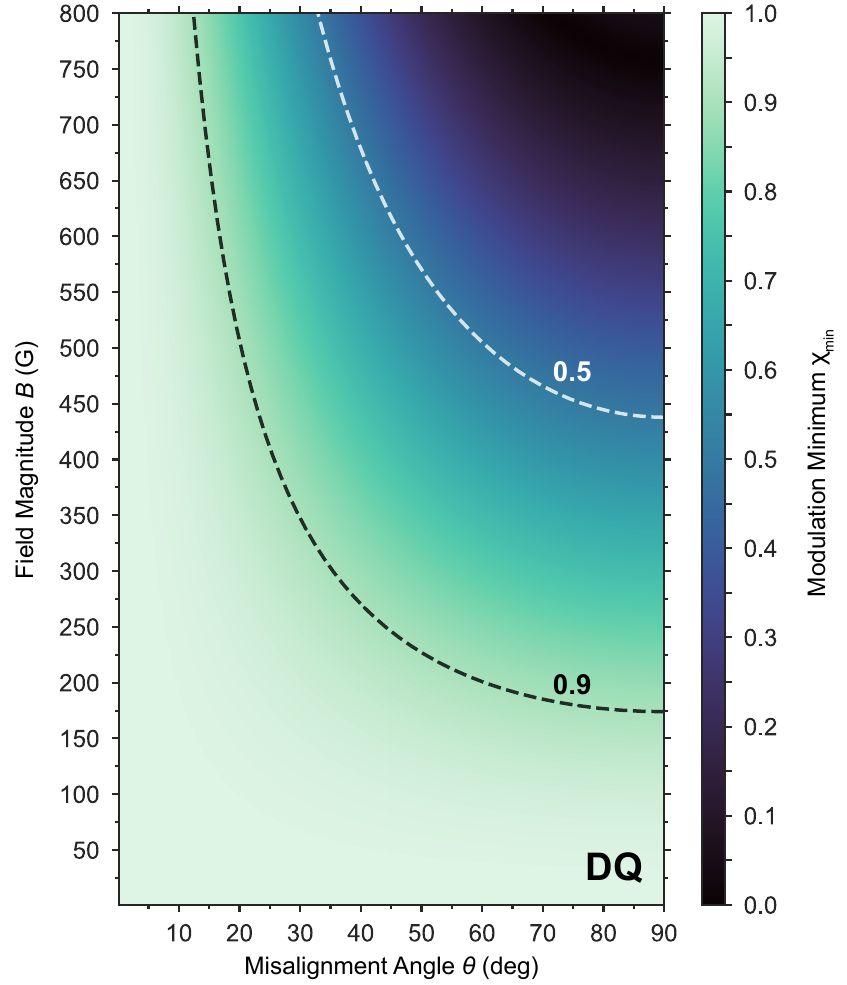}
    \caption{Calculated double-quantum (DQ) Ramsey relative contrast at amplitude modulation nodes $\chi_\text{min} = |\cos{(\Phi_{-1,+1})}|$, for magnetic field magnitudes $B$ up to 800\,G and misalignment angles up to $90^\circ$. Dashed contour lines indicate magnetic field configurations where $\chi_\text{min}$ equals 0.9 and 0.5, representing 10\% and 50\% contrast loss at envelope nodes, respectively.}
    \label{fig:DQ2D}
\end{figure}

In Fig. \ref{fig:5}(c) of the main text, 2D color plots of the relative contrast at amplitude modulation nodes $\chi_\text{min}$ are shown for both single-quantum (SQ) and double-quantum (DQ) Ramsey. The calculations for DQ Ramsey are reproduced here, with the range of magnetic field magnitudes and misalignment angles extended to 800\,G and $90^\circ$, respectively. For the magnetic field magnitudes considered in this study $B\lesssim200\,G$, the contrast at envelope modulation nodes is well-preserved ($\chi_\text{min}\approx1$). However, the calculations illustrated in Fig. \ref{fig:DQ2D} show that for a sufficiently large magnetic field magnitude, the DQ Ramsey signal can exhibit significant envelope modulations. 

\end{document}